# HOW TO USE QUANTUM THEORY LOCALLY TO EXPLAIN "NON-LOCAL" CORRELATIONS

Richard Healey[1]

**Abstract**
J.S. Bell's work has convinced many that correlations in violation of CHSH inequalities show that the world itself is non-local, and that there is an apparently essential conflict between any sharp formulation of quantum theory and relativity. Against this consensus, this paper argues that there is no conflict between quantum theory and relativity. Quantum theory itself helps us explain such (otherwise) puzzling correlations in a way that contradicts neither Bell's intuitive locality principle nor his local causality condition. The argument depends on understanding quantum theory along pragmatist lines, and on a more general view of how that theory helps us explain. Quantum theory is compatible with Bell's intuitive locality principle and with his local causality condition not because it conforms to them, but because they are simply inapplicable to quantum theory, as so understood.

## 1. Introduction

In a series of seminal papers, John Bell argued that the predictions of any acceptable theory satisfying a condition he called "local causality" must, in some cases, conflict with those of quantum theory. These cases concern certain correlations displayed by pairs of observable macroscopic events, where each event in a pair occurs at essentially the same time but in a different place. Bell concluded that quantum theory is not a locally causal theory. Subsequent experiments have amply confirmed quantum theory's predictions in these cases. Bell took his condition of local causality to be motivated by the intuitive principle that "The direct causes (and effects) of events are near by, and even the indirect causes (and effects) are no further away than permitted by the velocity of light" (Bell [1990], p.105), a principle closely associated with relativity theory. Quantum theory's failure to be locally causal puts it in serious tension with this intuitive principle and the relativity theory associated with it. But quantum theory's success in predicting the observed correlations then challenges the intuitive principle itself, and makes one reconsider its association with relativity.

The dialectic of the previous paragraph has produced something of a consensus among those most closely influenced by Bell's writings that the observed correlations demonstrate the falsity of the intuitive principle—in that sense the world itself is non-local. There is less agreement about the implications for relativity. Some maintain the need for a preferred frame in any acceptable theory capable of accounting for the observed correlations, while acknowledging that theory's implication that this frame is empirically inaccessible. Others continue to seek a relativistically invariant account of the observed correlations within a theory that is only approximately empirically equivalent to quantum theory. But again there is widespread agreement with Bell that there is an "apparently essential conflict between any sharp formulation [of quantum theory] and relativity." (Bell [2004], p.172)

[1] University of Arizona
Tucson, Arizona 85721-0027, USA
rhealey@email.arizona.edu

Against this consensus, I shall argue that there is no conflict between quantum theory and relativity, and that quantum theory itself helps us explain the (otherwise) puzzling correlations in a way that contradicts neither Bell's intuitive locality principle nor his local causality condition. The argument will depend on understanding quantum theory along pragmatist lines I have outlined elsewhere [in press], and on a more general view of how that theory helps us explain [forthcoming]. To avoid misunderstandings I begin by pointing out that quantum theory is compatible with Bell's intuitive principle and with his local causality condition not because it *conforms to* them, but because they are simply inapplicable to quantum theory, as so understood.

I foresee two reactions among readers less familiar with the grounds for consensus among those most closely influenced by Bell's writings. People continue to claim that the upshot of Bell's theorem in its various forms is that no local realistic theory can account for violations of so-called Bell inequalities. If you hold this opinion, then you may think the reason why quantum theory is able to offer a convincing local explanation of correlations that violate these inequalities is simply that it is not a realistic theory. There is a sense of "realism" in which something like this is right. But there are other senses in which it is wrong. Those making the claim rarely say exactly what they mean by "realistic", and even more rarely show how giving up realism enables us to use quantum theory satisfactorily to explain violations of Bell inequalities.

Other readers may see no need to interpret quantum theory one way rather than another in the light of the theory's successful prediction of "non-local" correlations, including violations of Bell inequalities. For what more could be required of an explanatory theory than that it enable one successfully to anticipate nature in such an impressive manner? They may go on to note the extraordinary success of relativistic quantum field theory in its Poincaré invariant form and to point to its satisfaction of the so-called microcausality condition (vanishing of commutators between local observables at space-like separation) as demonstrating conformity both to local causality and to the demands of relativity.

But matters are not so simple, as Bell's brilliant papers make clear. Bell himself anticipated this appeal to relativistic quantum field theory and convincingly explained why it fails to square quantum theory with local causality or fundamental Poincaré invariance. And his severe critique of contemporary formulations of quantum theory casts serious doubt on the adequacy of its explanations of the puzzling correlations associated with his name. In the next section of this paper I present Bell's criticisms, and in section 3 I show how a pragmatist interpretation of quantum theory is able to evade them.

Recent work by several authors[1] has further honed Bell's argument as to why any theory proposed as a candidate for explaining EPR-Bell correlations must violate local causality. This work is recalled in section 4. Section 5 then uses the pragmatist interpretation sketched in section 3 to explain why quantum theory itself is not a candidate theory in this sense, so the argument does not apply to it. With the ground thus cleared, section 6 shows how quantum theory helps explain "non-local" correlations involving two or more "particles", including those associated with the GHZ state.

While Bell himself allowed that the statement of probabilistic independence he derived from his locality condition may not embody everyone's idea of local causality, some have adapted independently developed analyses of causal relations to this case to argue that the failure of such probabilistic independence does indeed evince superluminal causation. In section 7 I address arguments for superluminal causation that seek to exploit connections between causal and counterfactual relations. While acknowledging some such connection, I reply that the only counterfactuals that hold in this case manifest epistemic rather than causal connections between distant events. Section 8 goes on to explore how quantum theory exploits the possibility of private informational links between an agent and events (s)he neither observes nor brings about, in ways that are strikingly independent of the spatiotemporal relations between them. This possibility has interesting implications for theories of chance in a relativistic world. In conclusion I distinguish various senses of 'non-local' and review how we are able to use quantum theory locally to explain "non-local" correlations in a relativistically invariant manner without recourse to any superluminal causation.

## 2. Bell's critique of current formulations of quantum mechanics: A response

It is ironic that the man who first realized the foundational significance of the fact that quantum mechanics predicts violations of his eponymous inequalities was also one of the severest critics of the present state of that theory. Bell, at least, did not accept that we have an exact formulation of a serious part of quantum theory capable of providing us with a satisfactory explanation of the violations of Bell inequalities it so successfully predicted. The best way to introduce a way of understanding the quantum theory that, I maintain, helps us locally to explain "non-local" correlations is to show how it evades Bell's critique of contemporary formulations of quantum theory.

Bell rejected as inexact any formulation of quantum theory that used words like 'macroscopic', 'irreversible', 'information' or (especially) 'measurement' in what he called its fundamental interpretive rules. He thought that the presence of such words in a formulation of quantum theory introduces an unacceptable imprecision that leaves entirely too much to the discretion of the theoretical physicist. He especially focused his critique on the role of 'measurement' in stating the Born Rule linking a quantum state to probabilities *of measurement outcomes*, and the effect of *measurement* in "collapsing" the quantum state onto an eigenstate of the measured observable. As he put it

> …the word comes loaded with meaning from everyday life, meaning which is entirely inappropriate in the quantum context. When it is said that something is 'measured' it is difficult not to think of the result as referring to some pre-existing property of the object in question. …
> 
> the word has had such a damaging effect on the discussion, that I think it should now be banned altogether in quantum mechanics. (Bell [2004], p.216)

By itself, this semantic objection is not convincing. Most, if not all, quantum "measurements" of an observable are processes that would count classically as revealing some pre-existing property—the value of the corresponding dynamical variable. Preservation of the reference of a term like 'measurement' under scientific change is standard linguistic practice. Secondly, what are called quantum "measurements" can provide vital information, which is the essential function of any measurement.

But Bell had a more substantial objection to the use of 'measurement' in a formulation of quantum theory. In his words:

> The first charge against 'measurement', in the fundamental axioms of quantum mechanics, is that it anchors there the shifty split of the world into 'system' and 'apparatus'. (Bell [2004], p.216)

The split (often called the Heisenberg cut or *schnitt*) is manifest in the different ways 'system' and

'apparatus' are treated in the Born Rule for single probabilities, which may be stated as follows

$$\text{prob}_\rho(Q\varepsilon\Delta) = \text{Tr}(\rho \mathbf{P}^Q[\Delta]) \qquad (1)$$

where $\rho$ is a density operator representing a quantum state and the statement '$Q\varepsilon\Delta$' locates the value of dynamical variable $Q$ (on some system $s$) in set $\Delta$ of real numbers.[2] The split appears between system (assigned state $\rho$) and apparatus (whose final condition is described as registering outcome $Q\varepsilon\Delta$). However this division is made, the Born Rule treats system and apparatus differently. The condition of the system is specified by assigning it a quantum state, while that of the apparatus is given by assigning it a property that serves to register the outcome.

Dirac and von Neumann proposed to paper over this division by taking the quantum state of a system indirectly to assign it corresponding properties, in accordance with what has become known as the (generalized) eigenvalue-eigenstate link:

(EE)  A system has property $Q\varepsilon\Delta$ if and only if its quantum state is an eigenstate of $\mathbf{P}^Q[\Delta]$ with eigenvalue 1.

(EE) implies that a system has a dynamical property in a quantum state just in case the Born Rule assigns the property probability 1 in that state.

This does not yet restore parity between system and apparatus. If the apparatus is itself a quantum system, then its condition should also be given by a quantum state that serves to register the outcome in accordance with (EE). If the system is initially in an eigenstate of $\mathbf{Q}$ with eigenvalue $q_i$ while the apparatus is (a quantum system) in an eigenstate of an operator $\mathbf{R}$ corresponding to recording dynamical variable $R$, then all is well. Formally, there is a unitary evolution

$$|q_i\rangle \otimes |r_0\rangle \to |\psi_i\rangle \otimes |r_i\rangle \qquad (2)$$

interpretable in accordance with (EE) as an interaction between a system $s$ on which $Q$ has value $q_i$ before the interaction (and afterwards just in case $|\psi_i\rangle = |q_i\rangle$), while the value of $R$ on an apparatus system $m$ changes from its initial value $r_0$ to a final value $r_i$ that does indeed serve to record on $m$ the value of $Q$ on $s$. But notoriously all is *not* then well if the system is initially in a non-trivial superposition of orthogonal eigenstates of $\mathbf{Q}$

$$(\textstyle\sum_i c_i |q_i\rangle) \otimes |r_0\rangle \to \sum_i c_i (|\psi_i\rangle \otimes |r_i\rangle) \qquad (3)$$

It now follows from (EE) that after the interaction $R$ on apparatus system $m$ has no value that can serve to register an outcome—in manifest contradiction with experiment.

Bell ([2004], chapter 23) documented two responses to this difficulty, one by Dirac, the other by Landau and Lifshitz. Dirac's response was more radical: he simply denied that the quantum state of a system evolves unitarily during a measurement on that system, and indeed refused to treat this as a quantum mechanical interaction between system and apparatus. Accordingly, he replaced (3) by the stochastic evolution

$$(\textstyle\sum_i c_i |q_i\rangle) \to |q_i\rangle \text{, with probability } |c_i|^2 \qquad (4)$$

(which von Neumann called process 1). Landau and Lifshitz endorsed (3), but then appealed to the special classical nature of the apparatus to infer (in apparent contradiction to (EE)) that after the measurement $R$ has some value $r_i$ on $m$, with probability $|c_i|^2$, and that the state of $s$ is $|\psi_i\rangle$.

This lack of unanimity among canonical formulations of quantum theory is already worrying. Moreover both responses apparently attribute to measurement a unique dynamical role in quantum theory, since it involves a stochastic change in the quantum state of the measured system quite different from deterministic, linear Schrödinger evolution—the notorious collapse of the wave-packet. If there are indeed two dynamic processes inherent in quantum theory, then a precise formulation of the theory must specify exactly when each process occurs. Bell inveighed against the lack of precision in formulations that took collapse to occur only on "measurement", interaction with a "classical" or "macroscopic" system, in an "irreversible" interaction, etc. He pointed out that simply by reconsidering the division between quantum system and classical apparatus a physicist can change the predictions of the theory.

> the following rule for placing the Heisenberg split, although ambiguous in principle, is sufficiently unambiguous for practical purposes:
>
> *Put sufficiently much into the quantum system that the inclusion of more would not significantly alter practical predictions.*
>
> .....
>
> The problem is this: quantum mechanics is fundamentally about 'observations'. It necessarily divides the world into two parts, a part which is observed and a part which does the observing. The results depend in detail on just how this division is made, but no definite prescription for it is given. All that we have is a recipe which, because of practical human limitations, is sufficiently unambiguous for practical purposes.
>
> (Bell [2004], p.124)

We can distinguish two separate problems here. If collapse is a physical process that discontinuously changes a system's quantum state on measurement, and that state completely describes the system's dynamical properties in accordance with (EE), then what properties quantum theory ascribes to a system at a time depends on whether a quantum state of (a super-system of) the system has collapsed by that time—and that depends on the placement of the division between system and apparatus. This is an internal consistency problem that arises only once one accepts (EE) and that wave-collapse is a physical process. The second problem is that the "practical predictions" of quantum theory are themselves sensitive to the placement of the Heisenberg cut. That is, by choosing to move the cut a physicist could alter what quantum theory predicts would be observed in a given situation, however difficult it may be experimentally to realize that situation and to perform observations capable of discriminating between these alternative predictions.

Bell admitted that this second problem did not affect the adequacy of quantum theory for all practical purposes (FAPP). To argue that FAPP is not good enough he compared quantum mechanics unfavorably to classical mechanics:

> In classical mechanics we have a model of a theory which is not intrinsically inexact, for it neither needs nor is embarrassed by an observer.
>
> At least one can envisage an accurate theory, of the universe, to which the restricted account is an approximation.
>
> (Bell [2004], p.125)
>
> …[quantum] theory is fundamentally approximate.
>
> …quantum mechanics…is…intrinsically inexact.
>
> (*op. cit.*, p.126, p.127)

I respond by first pointing out that *every* application of a physical theory, whether classical or quantum, is FAPP. To apply a theory one must consider some physical system and either ignore the rest of the world or model its effect on this system in a highly simplified way. In both classical and quantum theories, including more in the system may lead to more accurate (and *a fortiori* different) predictions at the cost of greater complexity. This is true even if the system is the whole universe, as it often is in theories of cosmology—where no model is expected to include minor details of local happenings such as what the cosmologist had for breakfast.

Bell's reply would be that only in classical physics can one "envisage" a model representing the whole universe, in all its details. If the theory were true, this model could be perfectly accurate. Nothing in the model need be designated "observer": though some parts of the model might be supposed to represent observers as physical systems. Simplifications of this model might "in principle" be arrived at by controlled approximations.

But there are serious grounds for skepticism here. It may be incoherent to suppose that anyone could build *himself* (as a physical system) into the model and then go on to use it to predict details of his own behavior. In any case, such a complete, detailed and accurate model would exceed human cognitive limitations, in respect of data gathering, presentation, and processing. One might imagine an enormously complex mathematical structure intended to represent the universe completely accurately within classical physics. But only in use would this structure represent the universe or anything else, for representation is a three-place relation between structure, target *and user*. It would then follow from the impossibility of using a structure within classical physics rich enough to count as a complete, detailed and accurate model of the universe that one *cannot* envisage an accurate classical physical theory, of the universe, to which a restricted account is an approximation. To be accurate, the theory would have to permit a representation of the universe in all its details. But that would require us to be able to deploy a mathematical structure so complex as to exceed our human limitations. Needless to say, this is in no way an indictment of classical physics, which proved very successful in fulfilling all the essential scientific functions of predicting, controlling and explaining phenomena until its limitations in these respects were revealed and then exceeded by quantum theory.

Even though the second "problem" is therefore seen to dissolve under closer examination, Bell's analysis correctly exhibited a disanalogy between quantum and classical physics. In classical physics various elements of a model (dynamical variables, system state and trajectory in phase space, etc.) are used to represent (a portion of) the physical world when the model is applied. But in quantum theory as usually understood, it is a central function of corresponding elements of a model (observables, quantum state and evolution in Hilbert space) to issue in predictions for the statistics of observations on something in the physical world. They can serve this function only if whatever plays the role of observer does not itself correspond to anything in the model. So in quantum theory there is an additional reason why one cannot envisage a complete, detailed and accurate model of the universe. This disanalogy is interesting, and provides a clue to a better way of understanding quantum theory. But it does not reveal any *defect* in quantum theory as currently formulated, as Bell's charges that the theory is fundamentally approximate and intrinsically inexact seem to imply.

What about the first problem raised by Bell's discussion of the shifty split? The problem was that what properties quantum theory ascribes to a system at a time depends on whether a quantum state of (a super-system of) the system has collapsed by that time—which in turn depends on the placement of the division between system and apparatus. As I noted, this problem arises only on the assumptions that collapse is a physical process which discontinuously changes a system's quantum state on measurement, and that a system's state completely describes its dynamical properties in accordance with (EE). But these assumptions are not clearly part of orthodox quantum theory, despite figuring in Dirac's and von Neumann's seminal texts.[3] The cleanest way to solve this problem is simply to drop both assumptions and to articulate an interpretation of quantum theory without them. I turn now to that task.

## 3. Sketch of a pragmatist interpretation of quantum theory

According to the pragmatist interpretation of quantum theory I outline in ([in press]), neither quantum states nor observables nor dynamical laws represent or describe the condition or behavior of any physical system. These elements function in other ways when an agent uses a quantum model to predict or explain phenomena by applying quantum theory. (EE) fails not because the quantum state incompletely describes an individual physical system (as Einstein argued), but because it does not describe an individual physical system at all: the following weaker principle also fails.

*Eigenstate-Eigenvalue Implication (EVI)*  If a system's quantum state is an eigenstate of an observable, then that observable has the associated eigenvalue.

This pragmatist interpretation rejects *EVI* and assigns the quantum state two roles. One is in the algorithm provided by the Born Rule for assigning quantum probabilities to appropriate claims of the form $Q_\Delta(s)$ : *The value of Q on s lies in Δ,* where $Q$ is a dynamical variable, $s$ is a quantum system and $\Delta$ is a Borel set of real numbers. But the significance of a claim $Q_\Delta(s)$ varies with the circumstances to which it relates. Accordingly, a quantum state plays a second role by modulating the *content* of a claim $Q_\Delta(s)$ by modifying its inferential relations to other claims.

Any application of quantum theory involves claims describing a physical situation. While it is considered appropriate to make claims about where individual particles are detected contributing to the interference pattern in a contemporary two-slit interference experiment, claims about through which slit each particle went are typically alleged to be "meaningless". In its second role the quantum state offers guidance on the inferential powers, and hence the appropriateness, of descriptive claims of the form $Q_\Delta(s)$—of what I call a **N**on-**Q**uantum **M**agnitude **C**laim (NQMC).

The key idea here is that even assuming unitary evolution of the joint quantum state of system and environment, delocalization of system state coherence into the environment will typically render appropriate descriptive claims about experimental outcomes and the condition of apparatus and other macroscopic objects by endowing these claims with enough content to license an agent to adopt epistemic attitudes toward them, and in particular to apply the Born Rule. But an application of quantum theory to determine whether this is so will not require referring to any system as "macroscopic", as an "apparatus" or as an "environment". All that counts is how a quantum state of a super-system evolves under the influence of a Hamiltonian associated with an interaction between the system of interest and the rest of that super-system.

It is important to note that since the formulation of the Born Rule now involves no explicit or implicit reference to "measurement", Bell's strictures against the presence of the term 'measurement' in a precise formulation of quantum theory are met. None of the other proscribed terms 'classical', 'macroscopic', 'irreversible', or 'information' appear in its stead.

Since an agent's assignment of a quantum state does not serve to represent a system's properties, her reassignment of a "collapsed" state on gaining new information represents no change in that system's properties. That is why collapse is not a dynamic process, on this interpretation of quantum theory. Nor, of course, is Schrödinger evolution of a quantum state. A quantum model does not serve to represent change in dynamical properties of a system, because it does not represent the dynamical properties of any system. Consequently, a formulation of quantum theory has no need to include a statement distinguishing the circumstances in which different dynamical processes occur. The importance of eliminating this need should be clear from section 2's discussion of Bell's critique of orthodox formulations that use 'measurement' or other proscribed terms when drawing this distinction. An agent can use quantum theory to track changes of the dynamical properties of a system by noting what descriptive claims of the form $Q_\Delta(s)$ are both sufficiently licensed and warranted at various times. But quantum theory itself does not imply any such claim, even when an agent has correctly assigned a system a quantum state that assigns the claim probability 1 *via* the Born Rule.

Quantum states are relational on this interpretation. The function of Born probabilities is to offer an agent authoritative advice on how to apportion degrees of belief concerning appropriate claims of the form $Q_\Delta(s)$ which the agent is not currently in a position to check. It follows that a system does not have a unique quantum state. For when agents (actually or potentially) occupy relevantly different physical situations they should assign different quantum states to one and the same system, even though these different quantum states are perfectly objective. Each will consequently assign different Born probabilities to a single claim of the form $Q_\Delta(s)$ concerning system *s* in a given situation. These different probabilities will then be equally objective and equally correct. This feature of the interpretation will prove crucial in what follows.

On this pragmatist interpretation, quantum theory by itself can explain nothing, because it describes nothing. The novel elements appearing in its models represent no physical systems and cannot be used to demonstrate representations of the phenomena to be explained: For quantum theory does not imply statements one can use to make claims about natural phenomena that describe or represent features of those phenomena. But quantum theory nevertheless helps us to explain an extraordinary variety of regularities in the physical world using representational resources from outside of quantum theory. It can do this because there is more to the informational structures quantum theory supplies than theoretical models involving quantum states, Hamiltonian, Lagrangian and other operators, and (solutions to) the Schrödinger equation and relativistic generalizations. Implicit in the theory are rules for *using* these models to guide one in making claims and forming beliefs about physical systems to which such models may be applied *but which these models do not themselves describe or represent*.

The first step in using quantum theory to help explain a regularity such as those exhibited by EPR-Bell correlations is to say what the regularity is. The explanandum regularity itself must be expressed in suitable claims of the form $Q_\Delta(s)$ or other non-quantum claims taken to supervene on them, but the circumstances in which it obtains may be described in other non-quantum terms. Cartwright ([1983], chapter 7) called this first stage of theory entry giving a prepared description of the explanandum and the conditions under which it holds. But note that at this stage we have not yet entered the domain of *quantum* theory, since the prepared description is not in terms of quantum states and operators and there has been no mention of probabilities. When the mathematical modeling apparatus provided by quantum theory is now deployed, the prepared description of the regularity is *not* shown to be a deterministic or stochastic consequence of laws or principles of quantum theory, dynamical or otherwise.

Quantum theory plays a different explanatory role. This is to show that an agent should expect the regularity to hold under these conditions by applying one or more mathematical models of quantum theory to the prepared description. Quantum theory tells an agent what to expect by guiding his credences in claims of the form $Q_\Delta(s)$ that have now been taken to represent the explanandum regularity.

The Born Rule plays a key role here: it figures, explicitly or implicitly, in all explanatory applications of quantum theory. Quantum theory contributes to our explanatory projects by providing us with a general set of techniques for calculating Born probabilities that tell us what we should expect, in familiar as well as unfamiliar situations. These include general techniques for determining quantum states, since the Born Rule can play its role only when supplied with a quantum state. How these techniques guide an agent in applying quantum theory to explain "non-local" correlations will be a matter of some concern in section 6.

The Born Rule may be legitimately applied to determine probabilities only of those non-quantum claims of the form $Q_\Delta(s)$ that are suitably licensed in the conditions in which the relevant systems are taken to be present. When the Born Rule is used in explaining a regularity, it is the prepared description of the explanandum that specifies the relevant systems and surrounding conditions. Quantum theory furnishes no precise rule specifying exactly when this description legitimizes application of the Born Rule to non-quantum claims expressing the regularity to be explained. Knowing when it is legitimate to apply the Born Rule here is a learned skill on a par with those needed at an earlier stage in preparing the description of the explanandum regularity and assigning a quantum state to a selected system. One who lacks such skills cannot be said to know quantum theory, no matter how effectively he can write down and solve the Schrödinger equation and calculate Born probabilities from its solutions.

But the theory of decoherence can help to justify a decision to apply the Born Rule under conditions specified by the prepared description of an explanandum regularity. For one can appeal to that description in modeling the effect of environmental interactions on the target quantum systems the description takes to manifest the regularity, as a process by which the coherence of their quantum state is rapidly delocalized into the environment. If such decoherence selects a robust set of approximately orthogonal quantum states of target systems, where each of these states is correlated with a particular quantum state of the environment (or states of its subsystems), then expectations based on application of the Born Rule to the quantum state of the target systems will be reliably borne out.

In this section I have quickly sketched a pragmatist interpretation of quantum theory outlined more fully elsewhere ([in press]) in order to show how contemporary quantum theory, so understood, evades Bell's critique. I then showed in general terms how quantum theory helps us explain the kinds of regularities it is usually taken to explain: for a more complete discussion, with examples, see my ([forthcoming]). In the rest of this paper I focus on the explanation of "non-local" correlations, beginning with Bell's seminal argument that any theory proposed as a candidate for explaining EPR-Bell correlations must violate a condition of local causality.

## 4. The argument as to why candidate theories violate local causality

Bell's papers contain several formulations of his local causality condition, and arguments that the predictions of any theory meeting a condition so formulated must conform to an inequality violated in some cases by those of quantum theory. Perhaps his most careful formulations were in Bell [1990]. Norsen ([2009], [2011]) has sought further to clarify these arguments, and Seevinck and Uffink [2010] (*SU*) recently presented a persuasive reconstruction of Bell's argument that no locally causal theory can account for the patterns of statistical correlation expected on the basis of quantum theory and now amply confirmed by experiments. *SU* follow Norsen in stressing that Bell's condition of local causality is intended to apply to *theories* advanced as candidates for accounting for EPR-Bell correlations.

Bell's ([1990], p.106) formulation is as follows:

*Local Causality*  A theory is said to be locally causal if the probabilities attached to values of local beables in a space-time region 1 are unaltered by specification of values of local beables in a space-like separated region 2, when what happens in the backward light cone of 1 is already sufficiently specified, for example by a full specification of all local beables in a space-time region 3. The regions 1-3 are as indicated in figure 1.

*SU* first argue that it is not appropriate here to require a *full* specification of all local beables in region 3, since that would fix the apparatus setting in a Bell experiment with outcome in region 1, while this must be regarded as a free variable in deriving a Bell inequality on the basis of local causality (see Bell [1990], p.109)). Instead, *SU* set out to make more precise the notion of sufficiency that is important in deriving a mathematically sharp and clean version of local causality from the condition as formulated (except for the misleading example).

They clarify this notion of sufficiency as a combination of functional and statistical sufficiency, rendering the label $b$ and random variable $B$ (respectively) redundant for predicting $P_{a,b}(A|B,\lambda)$ — the probability a theory specifies for beable $A$ representing the outcome recorded in region 1 given beables $a,b$ representing the free choices of what the apparatus settings are in sub-regions of 1,2 respectively, conditional on outcome $B$ in region 2 and beable specification $\lambda$ in a region 3 which smoothly joins $3_a$ to a similar slice $3_b$ right across the backward light cone of 2 so as to screen off both 1 and 2 from the overlap of their backward light cones. (See figure 2.) This implies

$$P_{a,b}(A|B,\lambda) = P_a(A|\lambda) \quad\quad (5a)$$

By symmetry, interchanging '1' with '2', '$A$' with '$B$' and '$a$' with '$b$' implies

$$P_{a,b}(B|A,\lambda) = P_b(B|\lambda) \quad\quad (5b)$$

*SU* offer equations (5*a*) and (5*b*) as their mathematically sharp and clean (re)formulation of the condition of local causality. Together, these equations imply the condition

$$P_{a,b}(A,B|\lambda) = P_a(A|\lambda) \times P_b(B|\lambda) \quad\quad (6)$$

used to derive CHSH inequalities. Experimental evidence that these inequalities are violated just as quantum theory leads one to expect is then taken to disconfirm Bell's intuitive principle stated in section 1 above, and to highlight his opinion that there is an essential conflict between any sharp formulation of quantum theory and relativity.

*SU* endorse Bell's claim that orthodox quantum mechanics is not a locally causal theory, because it violates (6). In the Bell state $\Phi^+$, for example, the probability of recording a horizontally polarized photon in 1 depends on whether that polarization is recorded for the entangled photon in 2, since these records are perfectly correlated. The argument is that orthodox quantum theory specifies no beables in region 3 sufficient to render these outcomes probabilistically independent, as (6) requires. *EVI* would imply that quantum theory specifies *some* beables in region 3 for a system in the Bell state $\Phi^+$: the photon pair has a property associated with linear polarization, with respect to each axis in the relevant plane (*either* both along *or* both orthogonal, though *determinately* neither). But (6) fails even if one accepts *EVI* and takes $\lambda$ to be specified by these properties.

In more detail, *SU* claim that orthodox quantum mechanics violates the statistical sufficiency conditions

$$P_{a,b}(A|B,\lambda) = P_{a,b}(A|\lambda) \quad (7a) \quad\quad\quad P_{a,b}(B|A,\lambda) = P_{a,b}(B|\lambda) \quad (7b)$$

while conforming to the functional sufficiency conditions

$$P_{a,b}(A|\lambda) = P_a(A|\lambda) \quad (8a) \quad\quad\quad P_{a,b}(B|\lambda) = P_b(B|\lambda) \quad (8b)$$

Statistical sufficiency is a condition employed by statisticians in situations where considerations of locality and causality simply don't arise. But in this application the failure of quantum theory to provide a specification of beables in region 3 such that the outcome $B$ is always redundant for the probability of determining outcome *A* (and similarly with '*A*', '*B*' interchanged) has clear connections to local causality, as *SU*'s analysis has shown.

*SU* interpret the failure of (5*a*), (5*b*) in any candidate theory compatible with the quantum mechanical predictions for violation of CHSH inequalities as establishing a violation of local causality:

> there seems to be only one option, namely that local causality is violated, i.e. there must be some non-local causation present in the candidate theory under study. (p.13)

If quantum theory is a candidate theory, it would then follow that some non-local causation is present in quantum theory. In section 5 I will argue that quantum theory is *not* a candidate theory in this sense, and that quantum theory does not violate (5*a*), (5*b*) or (7*a*), (7*b*). These conclusions are directly connected. None of (5*a*), (5*b*) or (7*a*), (7*b*) are *well-defined* within quantum theory, precisely because it is not a candidate theory in the sense of Bell, Norsen and *SU*. Since they are not well-defined, they are not violated by quantum theory—but neither do they hold. (5*a*) and (5*b*) are mathematically sharp and clean (re)formulations of local causality when applied to a class of theories. But since quantum theory lies outside that class we can draw no conclusions about non-local causation in quantum theory from the work of Bell, Norsen and *SU*. I shall return to that issue in section 7 only after showing in section 6 how quantum theory helps us explain "non-local" correlations.

**5. Quantum theory is not a candidate theory**
Bell himself was quite explicit that his condition of local causality was designed for application to physical *theories* rather than to the world. He had two reasons for focusing on theories. First, it is easier to explore the causal structure of a theory than of the world:

> I would insist here on the distinction between analyzing various physical theories, on the one hand, and philosophising about the unique real world on the other hand. In this matter of causality it is a great inconvenience that the real world is given to us once only. We cannot know what would have happened if something had been different. We cannot repeat an experiment changing just one variable...Physical theories are more amenable in this respect. We can *calculate* the consequences of changing free elements in a theory, be they only initial conditions, and so can explore the causal structure of the theory.
> (Bell [2004], p.101)

Secondly, even our best theories remain a resource for saying something useful and possibly even true about the world: we cannot simply assume some theory under study adequately captures the nature of reality.

> The beables of the theory are those elements which might correspond to elements of reality, to things which exist. Their existence does not depend on 'observation'. Indeed observation and observers must be made out of beables. I use the term 'beable' rather than some more committed term like 'being' or 'beer' to recall the essentially tentative nature of any physical theory. Such a theory is at best a *candidate* for the description of nature. (Bell [2004], p.174)

To conclude from observed violations of Bell inequalities that *nature* violates local causality one would need reasons to believe that some such candidate theory truly describes the world: Though if the class of candidate theories is wide enough to include any plausible contender, then even the conclusion that there must be some non-local causation in a candidate theory would provide a reason to believe the world is non-local. So it is important to enquire into the qualifications for candidacy, and especially to ask whether any *prima facie* plausible contenders face premature disqualification. Bell's critique notwithstanding, quantum theory does enjoy the benefits of incumbency as our currently most fundamental physical theory. But I shall argue that these do not suffice even to qualify it for candidacy in this rigged election.

The argument is very simple. It is central to the pragmatist interpretation sketched in section 3 that quantum theory itself neither contains nor implies any descriptive claim about a physical system—neither about the world nor about any part of it. In Bell's terminology, quantum theory has no beables. What the theory does is to offer advice to an agent on the content and credibility of claims of the form $Q_\Delta(s)$. While such a claim is not part of quantum theory, an agent can certainly use it to describe system $s$. To see whether this makes $Q$ a beable we need to look more closely at what Bell means by this term.

Bell introduced the term 'beable' in contrast to quantum theory's 'observable':

> It is not easy to identify precisely which physical processes are to be given the status of "observations" and which are to be relegated to the limbo between one observation and another. So it could be hoped that some increase in precision might be possible by concentration on the *be*ables, which can be described "in classical terms", because they are there. The beables must include the settings of switches and knobs on experimental equipment, the currents in coils, and the readings of instruments. "Observables" must be *made*, somehow, out of beables. (Bell [2004], p.52)

He distinguished a class of local beables:
> We will be particularly concerned with *local* beables, those which (unlike the total energy) can be assigned to some bounded space-time region. ...
> It is in terms of local beables that we can hope to formulate some notion of local causality. (Bell [2004], p. 53)

And elsewhere:
> Now it may well be that there just *are* no local beables in the most serious theories. When space-time itself is "quantized", as is generally held to be necessary, the concept of locality becomes very obscure. ... So all our considerations are restricted to that level of approximation to serious theories in which space-time can be regarded as given, and localization becomes meaningful. Even then, we are frustrated by the vagueness of contemporary quantum mechanics. You will hunt in vain in the text-books for the local *be*ables of the theory. What you may find there are the so-called "local observables". It is then implicit that the apparatus of "observation", or better, of experimentation, and the experimental results, are real and localized. We will have to do as best we can with these rather ill-defined local beables, while hoping always for a more serious reformulation of quantum mechanics where the local beables are explicit and mathematical rather than implicit and vague. (Bell [1990], p.100)

Note that while Bell here contemplates the possibility of a "serious" theory of *non*-local beables, he (literally!) does not take seriously the possibility of a theory with *no* beables. A theory with no beables would not be a serious theory for Bell—i.e. a serious candidate for the job of truly describing nature. Note also that he takes for granted that claims describing experimental apparatus and the results of experiments are claims about beables, and that such claims should be founded on claims about the beables of a serious theory in two senses. First, the description offered by a serious theory should be sufficiently complete to determine the truth or falsity of any (sufficiently precise) description of the macroworld: and secondly, a serious theory should make deterministic or probabilistic claims that have testable implications for descriptions of the results of experiments.

So is $Q$ a beable in a claim of the form $Q_\Delta(s)$, about which an agent applying quantum theory may be advised to form a credence? Certainly an agent could make a claim of that form, and in so doing describe $s$, whether $s$ is macroscopic and observable or microscopic or otherwise unobservable. But there is reason to question whether this makes $Q$ or "the settings of switches and knobs on experimental equipment, the currents in coils, and the readings of instruments" beables. Bell took it that by assigning a (value of a) beable to an object one is simply assigning it an objective property: *be*ables, he said, can be described "in classical terms", because they are there.

Recall that, on the pragmatist interpretation of quantum theory sketched in section 3, the significance of a claim $Q_\Delta(s)$ varies with the circumstances to which it relates.[4] Accordingly, a quantum state plays a second role by modulating the content of a claim $Q_\Delta(s)$ by modifying its inferential relations to other claims. If $s$ is not the entire universe, then it is a subsystem of some larger system $v=s\oplus u$. Even if a claim $Q_\Delta(s)$ is licensed at time $t_1$ because of the extensive delocalization of coherence of the quantum state assigned to $s$ into the environment including $u$, quantum theory itself does not exclude the possibility of this decoherence spontaneously *re*localizing into a future quantum state of $s$ at $t_2$. One would ordinarily take it to follow from the claim $Q_\Delta(s)$ made at time $t_1$ that in either $s$ or its environment there must be traces at $t_2$ of the "objective property" of $s$ that $Q$ had a value in $\Delta$ at $t_1$. But even when an agent is both licensed by quantum theory and warranted by his experience to claim $Q_\Delta(s)$ at time $t_1$, he must acknowledge the (remote) possibility that at $t_2$ there will be absolutely no evidence that the value of $Q$ on $s$ lay in $\Delta$ at $t_1$, either in his own memory or in physical traces anywhere in the universe.

By modulating the *content* of a claim $Q_\Delta(s)$ attributing a value to a magnitude in this and other ways, quantum theory arguably deprives that magnitude of beable status. A beable was supposed to be something capable of corresponding to reality—such correspondence being what it takes to make a claim $Q_\Delta(s)$ true. But it is important to note that even if correspondence to reality is *not* its function, such a claim may still be accepted as true or false on the basis of observation and experiment, and so serve as evidence for (or against!) the quantum theory that advises an agent on its tenability.

Quantum theory is not a candidate theory, in Bell's sense. It posits no beables itself, and it even goes some way toward depriving claims external to quantum theory and made in the familiar language of everyday affairs, including laboratory procedures, of beable status. Bell would not count quantum theory, so interpreted, as a serious theory. But why not? Quantum theory understood along the lines of section 3 gives us our best and only way of predicting and explaining a host of otherwise puzzling phenomena. In particular, by its help we can explain "non-local" correlations including violations of Bell inequalities as well as results of experiments performed on GHZ states. The next section shows how.

## 6. How to use quantum theory locally to explain "non-local" correlations
In understanding how quantum theory contributes to an explanation of "non-local" correlations it is important to recall that quantum states are relational on this interpretation. When agents (actually or potentially) occupy relevantly different physical situations they should assign different quantum states to one and the same system. Each of these different quantum states is perfectly objective, and all may be equally correct: a system does not have a unique quantum state. When the Born Rule is used to calculate Born probabilities for suitably licensed claims of the form $Q_\Delta(s)$ concerning a system *s* in given circumstances, the relevant quantum state is that which pertains to the physical situation of an (actual or potential) agent to whom these probabilities offer authoritative advice on how to apportion degrees of belief concerning these claims in the agent's situation. It follows that sometimes differently situated agents should adopt different degrees of belief in a single claim of the form $Q_\Delta(s)$ concerning a system *s* in given circumstances. The different Born probabilities advising these different degrees of belief will then be equally objective and equally correct. None of them gives the *real* objective chance of $Q_\Delta(s)$ in the circumstances in which *s* finds itself, for there is no such thing.

So quantum theory should not be thought of as the kind of stochastic theory that specifies a unique chance to each possible outcome of a process in circumstances where that outcome is objectively undetermined by what has come before (however that is understood in a relativistic world). This is an important additional reason why quantum theory is not a candidate theory in the sense discussed in section 5. I noted in that section a reason why one who accepts quantum theory thereby accepts that a non-quantum claim of the form $Q_\Delta(s)$ is not a straightforward statement about a local beable, in Bell's sense. But there is a much stronger reason why an assignment of Born probability to such a claim is not a statement about a local beable: Born probabilities are certainly not local beables representing localized chances, if quantum theory is interpreted along the lines of section 3. Instead, they offer authoritative (different) advice to differently situated agents on what to expect, and thereby explain the statistical patterns each records.[5]

Consider figure 3, which depicts space-like separated measurements by Alice and Bob in regions 1,2 respectively on a photon pair in the Bell state $\Phi^+$. At $t_1$ each takes the polarization state of the *l-r* photon pair to be $|\Phi^+\rangle = 1/\sqrt{2}\,(|HH\rangle + |VV\rangle)$. At $t_2$, after recording polarization *B* for *r*, Bob ascribes state $|B\rangle$ to *l*, and uses the Born Rule to calculate $P(A) = |\langle A|B\rangle|^2$ for Alice to record polarization of *l* along the *a*-axis. At $t_2$, Alice ascribes state $\rho = \tfrac{1}{2}\mathbf{I}$ to *l*, and uses the Born Rule to calculate $P(A) = \tfrac{1}{2}$ that she will record polarization of *l* along the *a*-axis. Each wisely uses the calculated probability to guide his or her expectations as to the outcome of Alice's measurement. The question as to which, if either, of these different probabilities gives the real "chance" of Alice's outcome simply doesn't arise even though neither Bob's nor Alice's Born probability is subjective.

To repeat, Born probabilities aren't local beables representing localized chances: they offer authoritative (different) advice to differently situated agents on what to expect, and thereby explain the statistical patterns each records. Quantum theory does not prescribe a single probability for Alice's (Bob's) measurement outcome in 1(2), but probabilities tailored to the local physical situation of each of them. Quantum theory is simply not a candidate for application of Bell's local causality condition, since (at a time such as $t_2$) it does not specify unique probabilities attached to values of local beables in a space-time region.

With this in mind, look again at the equation (5*a*), one of the pair *SU* offer as their mathematically sharp and clean (re)formulation of the condition of local causality.

$P_{a,b}(A|B,\lambda) = P_a(A|\lambda)$ (5*a*) $\qquad P_{a,b}(B|A,\lambda) = P_b(B|\lambda)$ (5*b*)

(Parallel considerations apply to (5*b*)). Bell and others understand each term flanking the equality sign in (5*a*) as purporting to represent a single local magnitude—the chance of *A* in region 1, conditional on the beables $\lambda$ the theory specifies for region 3. But so understood, (5*a*) simply can't be applied in quantum theory, since this theory denies there is any such magnitude. This is to reiterate the previous point that Bell's local causality condition is inapplicable to quantum theory, as interpreted in section 3. But there may seem to be a different way to understand (5*a*) as stating a condition that *is* applicable within quantum theory.

Assume for the moment that both Alice and Bob are correct to agree in assigning state $|\Phi^+\rangle$ to the photon pair in region 3 (see figure 2), and also that this may be regarded as quantum theory's complete specification of beables in that region. The term $P_a(A|\lambda)$ on the right of equation (5*a*) may then be taken to specify the Born probability assigned by Alice to a photon being recorded as passing her polarizer oriented along the *a* axis, relative to the situation at space-time location marked by an X and labeled Alice-at-$t_2$ in figure 3 (or indeed relative to the situation at any space-time point in the backward light cone of 1). While the term $P_{a,b}(A|B,\lambda)$ on the left of equation (5*a*) may then be taken to specify the Born probability assigned by Bob to a photon being recorded as passing Alice's polarizer oriented along the *a* axis, relative to the situation at space-time location marked by an X and labeled Bob-at-$t_2$ in figure 3 (or indeed relative to any space-time point in the forward light cone of 2).

Reinterpreted in this way, it is clear why (5*a*) could not hold in quantum theory: the probabilities it then equates are relative to different physical situations, such that agents in the relevant physical situations (would) have access to different information. At $t_2$ Bob can know the outcome of his photon measurement, while Alice cannot. Similarly, the term $P_b(B|\lambda)$ on the right of equation (5*b*) is then reinterpreted as the Born probability assigned by Bob to a photon being recorded as passing his polarizer oriented along the *b* axis, relative to the situation at space-time location marked by an X and labeled Bob-at-$t_1$ in figure 3 (or indeed relative to the situation at any space-time point in the backward light cone of 2). The term $P_{a,b}(B|A,\lambda)$ on the left of equation (5*b*) is then reinterpreted as the Born probability assigned by Alice to a photon being recorded as passing Bob's polarizer oriented along the *b* axis, relative to the situation at space-time location marked by an X and labeled Alice-at-$t_3$ in figure 3 (or indeed relative to any space-time point in the forward light cone of 1). At $t_3$ Alice can know the outcome of her photon measurement, while Bob cannot.[6]

However, even though Alice and Bob may well be correct to agree in assigning state $|\Phi^+\rangle$ to the photon pair in region 3, in quantum theory this does *not* constitute an assignment of beables to that region: quantum theory has no beables, as presently interpreted! Quantum theory helps an agent such as Alice or Bob explain EPR-Bell correlations not by exhibiting them as outcomes of a deterministic or stochastic process involving beables (local or non-local), but by showing that these correlations are just what the agent should have expected in the circumstances. The agent does not use quantum theory itself in specifying these circumstances, but appeals to the theory only after the specification has been given in non-quantum terms. Moreover, strictly speaking, the correlations themselves are neither predicted nor even described by quantum theory. Rather, the theory offers an agent good advice on what to expect them to be.

To use quantum theory to explain "non-local" correlations, an agent needs to assign a quantum state to some system or systems in order to decide whether to expect them to obtain. In the case considered so far in this section, what makes $|\Phi^+\rangle$ the correct state for Alice and Bob to assign to the photon pair they take to figure in the explanation of the EPR-Bell correlations displayed by their measurements of photon polarization? By now the relevant community of physicists shares a body of wisdom concerning the circumstances in which such a state may be produced and certified as such. Foremost among currently recognized techniques for producing such a Bell state is spontaneous parametric down conversion of laser light by a specially cut nonlinear crystal such as beta barium borate (BBO), followed by passage through suitable optical devices such as frequency filters and polarization-transformers. To call the process spontaneous is to say that pairs appear at random intervals, presumably as a result of a stochastic process.

While quantum theory itself helps to explain at least some aspects of this process, its success in producing pairs in state $|\Phi^+\rangle$ may be independently certified by quantum state tomography. This involves measurements of just the kinds of "non-local" correlations that Alice and Bob want to explain with the help of quantum theory, thereby threatening an apparent circularity. But the threat can be countered in several ways. The state $|\Phi^+\rangle$ may be certified by measuring correlations between outcomes of measurements of circular polarization and linear polarization along just a few combinations of axes $a,b$, and then appealed to in the explanation of a potential infinity of correlations in other polarization measurements. With sufficient experience in setting up laboratory equipment, one can become confident even without performing quantum state tomography that one has indeed created a source of photon pairs in state $|\Phi^+\rangle$. Note that in either case, the warrant for assigning state $|\Phi^+\rangle$ rests on evidence characterizing laboratory equipment and its output in non-quantum terms. The process of producing pairs in state $|\Phi^+\rangle$ may be considered analogous to a process of synthesizing a complex organic molecule. Here, too, initial synthesis may be very difficult, and independent proof of success hard to come by: but with widespread practical experience and technical improvements the process may become reliable and routine.

There are other ways for producing the state $|\Phi^+\rangle$ and other ways for Alice and Bob to warrant their shared belief that this is the state of a photon pair on which they intend to make polarization measurements. How do they then appeal to this state in explaining the correlations between their outcomes? Each of them can offer a two stage account. While the accounts coincide at the first stage, they differ at the second.

Application of the Born Rule to $|\Phi^+\rangle$ yields joint probabilities for claims ascribing linear polarization $A$ (rather than $A^\perp$) to $l$ and $B$ (rather than $B^\perp$) to $r$ if and only if such claims both have a sufficiently wide license. Prior to interaction with their respective detectors, no such claims are licensed sufficiently for application of the Born Rule. Interaction between the quantized modes of the electromagnetic field and the detectors rapidly delocalizes the coherence of the field polarization modes, thereby extending to any agent including Alice and Bob a wide license to claim some outcome is registered by a photon detector in each of 1, 2 placed to detect photons emerging with polarization either parallel or orthogonal to $a,b$ respectively. This justifies each of Alice and Bob in applying the Born Rule prior to 1, 2 to calculate a joint probability distribution $P_{a,b,|\Phi+\rangle}(A,B)$ for the four possible outcomes of detection, either parallel or orthogonal to $a,b$. This distribution applies to every coincident pair of detections in this state, so such joint detections are probabilistically independent. It follows that each of Alice and Bob should confidently expect that the relative frequencies of detection events in a long sequence will closely match these joint probabilities. So for Alice, Bob and any other hypothetical agent in a relevant physical situation, the EPR-Bell correlations displayed in these conditions are just what were to be expected on the basis of quantum theory. These correlations are thereby explained.[7]

The explanation exhibits the dependence of these "non-local" correlations on all the non-quantum claims agents like Alice and Bob were each warranted in making before being in a position to certify the occurrence of the correlated events. In this sense it shows how this regularity is determined by the objective prior state of the world. But this dependence differs from causal determination in significant respects. It is indirect, in so far as it is mediated by a quantum state $|\Phi^+\rangle$ that does not have the function of ascribing properties to the photon pairs involved. Both that state and the Born probabilities it yields are relational, depending on the physical situation of an agent ascribing that state (though the situations of agents like Alice and Bob at $t_1$ lead them to ascribe the same state). Most strikingly, this dependence does not conform to (6) when λ is specified by all non-quantum claims agents like Alice and Bob were warranted in making before being in a position to certify the occurrence of the correlated events—i.e. claims about what happens to the past of a space-like hypersurface to the past of both 1 and 2. So this is not a dependence on a common cause in any familiar sense.[8] Finally, even though they are objective, the non-quantum claims that warrant agents like Alice and Bob in assigning state $|\Phi^+\rangle$ are not exactly claims about beables, in Bell's sense, as section 5 explained.

At $t_2$ Bob can go on to offer an explanation of why Alice gets the outcome she does when measuring the polarization of the particular photon $l$ paired with the photon $r$ whose polarization he has recorded in 2. The case is particularly striking when the axes $a,b$ coincide, since Bob is then in a position to explain why Alice's outcome certainly matches his own: but an assignment of any specific probability has significant explanatory value. Bob's application of the Born Rule to calculate the probabilities of Alice's outcomes is warranted by decoherence at the detectors as before. But Bob is now in a position to assign quantum state $|B\rangle$ to $l$, since he has recorded outcome $B$ in 2 and the quantum state assigned to the pair prior to 2 was $|\Phi^+\rangle$. Bob is justified in "collapsing the wave" in this situation not because there has been any physical change in the quantum state—remember, on this interpretation a quantum state never physically changes since it does not represent the condition of any physical system. Nor does Bob's replacement of $|\Phi^+\rangle$ by $|B\rangle$ reflect the fact that the polarization of $l$ is $B$ after 2—neither Bob nor anyone else was warranted in ascribing any determinate polarization to $l$ at $t_2$, since $l$ undergoes delocalization of coherence only at Alice's detector. Bob replaces $|\Phi^+\rangle$ by $|B\rangle$ simply as a way of updating his authoritative source of advice concerning matters about which his present physical situation prevents him from otherwise obtaining information.

To reinforce this point notice that if Alice is moving uniformly toward Bob at a high enough speed 1 will occur earlier than 2 in her frame. In that case, she can offer an explanation of why Bob gets the outcome he does when measuring the polarization of the particular photon $r$ paired with the photon $l$ whose polarization she has recorded in 1. At $t_3$, Alice is then in a position to assign quantum state $|A\rangle$ to $r$, since she has recorded outcome $A$ in 1 and the quantum state assigned to the pair prior to 2 was $|\Phi^+\rangle$. Alice's replacement of $|\Phi^+\rangle$ by $|A\rangle$ in her situation then represents no physical collapse, and is perfectly consistent with Bob's replacement of $|\Phi^+\rangle$ by $|B\rangle$ in his situation.

Notice that, in these circumstances, while Bob can explain Alice's outcome by using quantum theory to show that it was just what he should have expected given his own outcome, Alice can explain Bob's outcome by using quantum theory to show that it was just what she should have expected given her own outcome. Each explanation exhibits the dependence of the event explained on features of the situation described by NQMC's. But the symmetry between Alice's and Bob's explanations makes it clear that this dependence seems asymmetric only when relativized to the situation of an actual or hypothetical agent. There is nothing here to suggest that the dependence is causal or metaphysical rather than epistemic in origin, a point that will be pursued in section 7.

One can get further insight into how an agent's assignment of quantum states depends on the agent's physical situation by reflecting on a scenario in which Alice's measurement of $a$-polarization on $l$ occurs time-like later than Bob's measurement of $b$-polarization on $r$, as depicted in figure 4. In region I of her world-line, Alice is space-like separated from Bob's measurement of *b*-polarization on *r*, and so inevitably ignorant that its outcome is *B*. The objectively correct quantum state for Alice to assign to *l* is therefore $\rho = \frac{1}{2}\mathbf{I}$. But in region II of her world-line, Alice is in a position to obtain information about Bob's outcome, whether or not she has the physical means to access that information. Consequently, the objectively correct quantum state for Alice in region II to assign to *l* is |B>, although if she then in fact has no information about Bob's outcome, she should continue to assign *l* the mixed state $\rho = \frac{1}{2}\mathbf{I}$ as a way of acknowledging that, given her actual information, Bob's outcome could equally well have been $B^\perp$.

Notice that Alice's reassignment of quantum state to *l* does not occur at the time when she takes Bob's measurement of *b*-polarization on *r* to occur, but later, as her world-line crosses into the forward light cone of that event. Notice also that this reassignment of quantum state to *l* reflects no physical event occurring either on Alice's world-line or anywhere in the forward light cone of Bob's measurement. Certainly it reflects no physical event involving *l*. This just reinforces the lesson that "wave collapse" is not a physical process.

The "non-local" correlations manifested by GHZ states are striking since the probabilities that figure in them are extremal, taking on values 0 or 1, so the correlations may be considered strict. Consider a scenario in which each of Alice, Bob and Charlie performs some polarization measurement on one of three photons in the entangled polarization state

$$|GHZ> = 1/\sqrt{2} \, ( \, |HHH> + |VVV>) \qquad (9)$$

Specifically, each has a choice to measure either whether that photon is left- or right-circularly polarized, or whether than photon is linearly polarized along or orthogonal to an axis H´ at 45° with respect to the original H axis. If any two of Alice, Bob and Charlie choose to measure circular polarization while the third chooses to measure linear polarization with respect to H´, then the Born Rule applied to state |GHZ> implies with probability 1 the following perfect correlation between their outcomes: if the circular polarization measurements agree, then the linear polarization measurement has outcome V´, while if they disagree then the linear polarization measurement has outcome H´. But if all three choose to measure linear polarization with respect to H´ then if any two of their outcomes agree the third outcome is sure to be H´, while if any two of their outcomes disagree then the third outcome is sure to be V´. In section 7 I examine the implications of these predictions, verified by Pan *et.al.* (2000), for locality. But how can one use quantum theory to explain them?

Suppose Alice has chosen to measure linear polarization and knows that both the others have chosen to measure circular polarization. Initially she assigns state |GHZ> to all three photons, and traces over her polarization space to assign completely mixed polarization state $1/4(\mathbf{I}\otimes\mathbf{I})$ to Bob and Charlie's photon pair. After recording outcome H´ (say) she reassigns entangled state $1/\sqrt{2}(|RL>+|LR>)$ to Bob and Charlie's photon pair. She can now use this to explain the perfect anticorrelations between the outcomes of Bob's and Charlie's measurements. If she enters the forward light cone of Bob's measurement event first, her correct assignment to the state of Charlie's photon will be either |R> or |L>, depending on whether Bob's outcome is L or R. Neither of these reassignments of quantum state reflects any change in the properties of Bob's or Charlie's photon pair or its elements. Each of Bob and Charlie has his own similar alternative reassignment of quantum states to the photon pair he does not measure, reflecting changes in his own space-time location and the information available to him from that location. The only physical events figuring in any of their accounts are the local polarization measurements on the three photons. Each can appeal to the outcome of his own measurement in using quantum theory to help explain the perfect correlations between the outcomes of the other two measurements.

     Notice that, depending on the outcome of her measurement, Alice can explain either the perfect anticorrelation between Bob and Charlie's outcomes, or their perfect correlation. For if she had obtained outcome V´ she should instead reassign entangled state $1/\sqrt{2}(|RR\rangle+|LL\rangle)$ to Bob and Charlie's photon pair. Alice's measurement does not *produce* either correlation or anticorrelation between Bob's and Charlie's photons. It just provides Alice with information relevant to understanding whether the outcomes of their measurements will be correlated or uncorrelated. Here as elsewhere assignment of an entangled quantum state to a compound system does not reflect any physical relation between its components. That is true of the GHZ state itself, which in the experiment of Pan *et.al*. (2000) was assigned on the basis of the outcome of a polarization measurement on a fourth "herald" photon produced from one of the two entangled pairs out of which the other three photons were selected after suitable optical processing.

## 7. Counterfactuals, causation and locality

"Non-local" correlations such as those displayed in EPR-Bell and GHZ scenarios are generally, though not universally, taken to support counterfactual claims.[9] The strict correlations predicted by applying the Born Rule to a Bell state such as $|\Phi^+\rangle$ provide a classic example of this. For any polarization axis *x*, if Alice and Bob each successfully measure the linear polarization with respect to *x* of a (different) photon in state $|\Phi^+\rangle$, the outcomes of their measurements will agree (with probability 1). But in so far as performance of a polarization measurement on a photon with respect to axis $x_1$ physically excludes simultaneous performance of a polarization measurement on that photon with respect to axis $x_2$ (if $x_2$ is neither parallel nor orthogonal to $x_1$), Alice and Bob can each successfully measure the linear polarization of their photon in state $|\Phi^+\rangle$ along at most one axis. Suppose Alice measures linear polarization along $x_1$ with outcome *o*. Then the perfect correlations predicted by state $|\Phi^+\rangle$ apparently warrant the following subjunctive conditional:

     If Bob were to measure linear polarization along $x_1$ he would get outcome *o*.

If Bob is assumed not to measure linear polarization along $x_1$ (whether or not he measures some other component of polarization—linear, circular or elliptical) this is a counterfactual conditional. The state $|\Phi^+\rangle$ further apparently warrants subjunctive conditionals with probabilistic consequents, such as

     If Bob were to measure linear polarization along an axis at 30º from $x_1$ he would get
     outcome *o* with probability 3/4.

Note that the warrant for such counterfactuals is independent of the space-time separation between events involved in Alice's measurement and in Bob's (hypothetical) measurement, including Alice's choice of what polarization component to measure (along with its physical implementation) and the outcome of that measurement (if performed). In particular, Alice's choice and outcome may be space-like separated from Bob's choice and outcome.

     Causal relations are intimately associated with counterfactual conditionals. Lewis ([1973], [1986]) offered an influential counterfactual analysis of causation. Butterfield [1992] argued that it is a consequence of this analysis that (if what measurement is performed at one wing has no effect on the outcome at the other wing then) outcomes at different wings of a Bell-EPR-type situation are directly causally connected, even when space-like separated. In a footnote he credited Clifton with strengthening this argument by considering the strict correlations that obtain in a GHZ scenario. Indeed, Clifton, Pagonis and Pitowsky (CPP) [1992] contains an argument that, according to Lewis, there is strict causal dependence (and therefore a direct causal relation)

between space-like separated events in that scenario.[10] I sketch this argument in terms of the strict GHZ correlations described in section 6.

Let ◯ denote a measurement of circular polarization with possible outcomes L and R, and let / denote a measurement of linear polarization with possible outcomes V´, H´. Then the following counterfactual is warranted by the strict GHZ correlations in state |GHZ>, where the first entry in the antecedent refers to Alice's (counterfactual) measurement, the second to Bob's, and the third to Charlie's, and similarly for the outcomes of the measurements in each disjunct of the consequent:

$$\{\bigcirc\bigcirc/\} \;\square\!\!\rightarrow\; \{(LLV´) \vee (RRV´) \vee (LRH´) \vee (RLH´)\} \qquad (10a)$$

Here $\square\!\!\rightarrow$ symbolizes the subjunctive conditional *if it were the case that....then it would be the case that* and ∨ symbolizes exclusive *or*. Two other counterfactuals (10b), (10c) are similarly warranted: these result from (10a) by permuting the measurements (and corresponding possible outcomes) among Alice, Bob and Charlie. So too is this fourth counterfactual:

$$\{///\} \;\square\!\!\rightarrow\; \{(V´V´H´) \vee (H´H´H´) \vee (V´H´V´) \vee (H´V´V´)\} \qquad (10d)$$

Now let λ specify beables in a region constituting a space-like slice right across the backward light cones of Alice, Bob and Charlie's (counterfactual) measurement events that screens these all off from the overlap of their backward light cones. Consistent with (10a) we must have

$$\{\bigcirc\bigcirc/\lambda\} \;\square\!\!\rightarrow\; \{(LLV´) \vee (RRV´) \vee (LRH´) \vee (RLH´)\} \qquad (11a)$$

and so (from the first and fourth disjuncts)

$$\{\bigcirc\bigcirc/\lambda\} \& \{L_a L_b\} \;\square\!\!\rightarrow\; V´_c \qquad (12ai)$$
$$\{\bigcirc\bigcirc/\lambda\} \& \{R_a L_b\} \;\square\!\!\rightarrow\; H´_c \qquad (12aii)$$

provided that the antecedent of each of these two counterfactuals holds in some physically possible world. But (12ai), (12aii) imply Lewisian strict causal dependence between Alice's outcome and Charlie's outcome, even if their measurements are space-like separated. If there is no such dependence, then at most one of equations (12a) has an antecedent that holds in some possible world. That in turn implies that at most one of the combinations $\{L_b, V´_c\}$, $\{L_b, H´_c\}$ can be compatible with $\{\bigcirc\bigcirc/\lambda\}$. Similar arguments based on the other five pairs of disjuncts of (11a) rule out other possible combinations of pairs of outcomes for two of the three measurements, consistent with $\{\bigcirc\bigcirc/\lambda\}$. So if there is no Lewisian causal dependence between outcomes for a particular λ, then that λ uniquely determines the outcomes of all three measurements $\{\bigcirc\bigcirc/\}$ in the GHZ state. Parallel arguments establish a similar conclusion based on each of the counterfactuals (11b)-(11d) that result from (10b)-(10d) by further specifying λ in the same way that (11a) resulted from (10a). So if there is no strict causal dependence between outcomes of Alice's, Bob's and Charlie's measurements, then those outcomes are determined by λ (with probability 1) in the GHZ state. Notice that the argument did not *assume* such determinism by λ, but proved it from a locality assumption (the absence of Lewisian strict causal dependence between outcomes for each particular λ). Notice also that the argument did not assume the following EPR sufficient condition for the reality of a physical quantity—a condition that has sometimes been identified with realism:

> If, without in any way disturbing a system, we can predict with certainty (i.e. with probability equal to unity) the value of a physical quantity, then there is an element of physical reality corresponding to this physical quantity. (EPR [1935], p.777)

Having used the GHZ correlations to establish determinism of measurement outcomes by λ, CPP's argument continues by using this to establish Lewisian causal dependence between measurement settings and distant outcomes. The first step is to show that if there is no such causal dependence then what the outcome of any of Alice, Bob or Charlie's polarization measurement is determined by λ to be cannot depend on what measurements the other two perform, on a given triple of photons. Now associate value +1(−1) with outcome R(L) of a circular polarization measurement, and value +1(−1) with outcome V´(H´) of a linear polarization measurement in the GHZ scenario. Then (10a)-(10c) imply that the product of the values associated with two circular polarization measurements and one linear polarization measurement is always +1, while (10d) implies that the product of the values associated with three linear polarization measurements is always −1. There is no assignment of predetermined outcomes to all these possible combinations of measurements compatible with these values. So no matter what λ may be, and no matter how it determines the outcomes of Alice, Bob and Charlie's measurements, no value of λ is compatible with the GHZ correlations predicted by quantum theory and now experimentally confirmed unless there is Lewisian causal dependence—either between some measurement setting and the outcome of a distant measurement, or between the outcomes of distant measurements.

When CPP come to consider how their argument bears on quantum theory, they conclude that it is determinism that fails, and so there is Lewisian strict causal dependence between distant outcomes in the GHZ scenario. They do not see this as raising an objection to Lewis's analysis, since such causal dependence cannot be used to construct tachyon-like "causal" paradoxes. Indeed, in their view Lewis's analysis is helpful in explicating the idea of peaceful coexistence between quantum theory and special relativity. It is worth noting that their discussion assumes that in quantum theory the GHZ state itself functions as a variable λ—an assumption rejected by the present pragmatist interpretation of that theory. But their sanguine endorsement of the view that GHZ correlations manifest superluminal causation should be rejected for a more important reason: The counterfactual connections between space-like separated events here are epistemic rather than causal—a position I shall defend after considering another attempt to use such counterfactual connections to argue for superluminal causation.

Without advancing a counterfactual theory of causation, Maudlin ([2011], p.118) defended a sufficient condition for causal connection of distinct events modeled on Bell's local causality condition.

> ...given a pair of space-like separated events *A* and *B*, if *A* would not have occurred had *B* not occurred even though everything in *A*'s past light cone was the same then there must be superluminal influences.

He applies this to an ideal experiment in which Alice and Bob have set their polarizers along the same axis *a* in the Bell state |Φ⁺> and their detectors record the same photon polarizations in space-like separated events *A*, *B*. Maudlin first argues that the antecedent of his sufficient condition is met by the events *A*, *B*, then applies this condition to conclude that there are causal connections between space-like separated events here—either directly between *A* and *B* or between at least one of these events and some common cause *C* that lies outside the overlap of their backward light cones. I shall argue that Maudlin's sufficient condition is incorrect. But first I question his argument seeking to establish the truth of its antecedent.

The statement *A would not have occurred had B not occurred even though everything in A's past light cone was the same* is a counterfactual. Maudlin wishes to show that it is true. While acknowledging the difficulties often encountered in evaluating counterfactuals, Maudlin claims that these can be overcome in certain scientific contexts:

Counterfactual claims can be implied by laws if the antecedent of the conditional is precisely enough specified. ([2011], p.120)

To use this to see if the key counterfactual used to formulate his sufficient condition for causal connection holds for quantum theory, it is then necessary to specify the laws of that theory. Maudlin takes the Schrödinger equation to be one such law stating how physical magnitudes will, or could, evolve through time. But the Schrödinger equation alone does not imply the relevant counterfactual: moreover, the present interpretation of quantum theory denies that it states how physical magnitudes will, or could, evolve through time. As Maudlin's discussion ([2011], pp.123-7) makes clear, the law that supposedly implies the key counterfactual can only be a stochastic law governing wave-collapse. Indeed, Maudlin's target in this whole argument is a common understanding of quantum theory, on which definite measurement outcomes are secured not by hidden variables but through a stochastic process grounding physical wave-collapse.

But, with no mention of hidden variables, the interpretation outlined in section 3 denies that quantum probabilities arise from the playing out of any such stochastic process. As presently interpreted, quantum theory involves no stochastic law governing wave-collapse: a quantum state does not describe a quantum system, and so an agent's reassignment of quantum state on gaining new information represents no physical change in a quantum system. So one cannot use laws of quantum theory to establish the truth of the counterfactual in the antecedent of Maudlin's sufficient condition.

Despite this failure, an agent like Bob-at-$t_2$ (in figure 3) who accepts quantum theory has good reason to endorse the counterfactual claim (*B*) in the case *a=b*.

(*B*) If the outcome of Bob's measurement had been different then the outcome of Alice's measurement would have been different, even though everything in the past light cone of 1 was the same.

For Bob is licensed to use state $|\Phi^+\rangle$ in the Born Rule to calculate the probabilities of Alice's two possible outcomes, conditional on his having obtained a different outcome. On the basis of that calculation Bob is wholly warranted in believing the counterfactual claim (*B*). Similarly, Alice-at-$t_3$ is wholly warranted in believing (*A*).

(*A*) If the outcome of Alice's measurement had been different then the outcome of Bob's measurement would have been different, even though everything in the past light cone of 2 was the same.

In this sense, Maudlin's sufficient condition for causal connection *can* be applied to quantum theory, though not on the basis of laws of that theory governing the evolution of any physical process. However it does not follow that the condition is correct, as so applied. I shall argue that it is *not* correct, as so applied. The counterfactual *A would not have occurred had B not occurred even though everything in A's past light cone was the same* is important to a situated agent not because it establishes a causal relation between space-like separated events, but because it establishes an *epistemic* link between them.

How does it serve Alice's and Bob's purposes to endorse the claims (*A*), (*B*) respectively? More generally, what role do counterfactuals play in an agent's practical reasoning? Since an agent can act only in the actual world, it is not immediately obvious why she should be concerned with counterfactual possibilities. The basic reason is that no agent can have complete information about the world in a situation in which she is deciding what to do. Since she must act on the basis of incomplete information it is important for her to consider different ways the world could be, consistent with whatever information she takes herself to have. It is also important for her to make the best use of the information available to her through theoretical reasoning. Such reasoning may be based on assumptions of which she is sure; it may be merely hypothetical—depending on assumptions she is not in a position currently to evaluate; or it may be based on alternative assumptions, any one of which she considers within her power to *make* true by her own free action.

This last use of theoretical reasoning provides one important role for counterfactuals in practical reasoning. In deciding what to do an agent needs to trace out the possible consequences of her alternative actions in order to evaluate them, as well as their relative probabilities. Therefore associated with each alternative possible action there is a variety of subjunctive conditionals concerning its possible consequences and some estimate of the probability of each consequence. Since only one alternative action can occur, most of these will have false antecedents, and be in this sense counterfactual. This is the home territory of *causal* counterfactuals. An agent needs causal knowledge in order to trace out the consequences of each of her alternative actions, and that is just what causal counterfactuals supply her with. Recent authors have noted how the advance of science has extended the use of causal counterfactuals far beyond this home territory to any situation in which it makes sense to speak of an *intervention* in the behavior of some system, where such intervention need not be the direct result of an agent's action.[11] Accordingly, if it does not make sense to speak of an intervention in a system capable of making the antecedent of a counterfactual conditional true, then this is not a causal counterfactual.[12] I shall argue that one *cannot* make sense of an intervention that secures the truth of the antecedent of (*A*) or of (*B*).

How could one try to make sense of this idea? Certainly no alternative action of Alice or Bob could make one of these true without disrupting the system involved in the EPR-Bell scenario itself (e.g. by preparing a different quantum state). In particular, choosing to measure a different polarization component would not have this effect. Quantum theory itself provides no resources on which one can draw to make sense of an intervention capable of changing the outcome of Alice's or Bob's measurement of a fixed component of polarization.

In his sophisticated discussion of what the possibility of intervention requires, Woodward ([2003], pp. 130-3) argues that an intervention must be conceptually possible, though it need not be physically possible. He considers a case in which an event *C* that is a potential locus of intervention occurs spontaneously in the sense that it has no causes. He argues that even in this case one can make sense of an intervention on *C*. This suggests that one can still make sense of the idea that an intervention in the EPR-Bell scenario is capable of making true the antecedent of (*A*) or of (*B*) in the case under consideration. But if one examines the conditions he imposes on an intervention it turns out that these cannot all be met here.

Woodward ([2003], p. 98) states necessary and sufficient conditions for *I* to be an intervention variable for *X* with respect to *Y*. These include

(I2)     *I* acts as a switch for all the other variables that cause *X*. That is, certain values of *I* are such that when *I* attains those values, *X* ceases to depend on the values of other variables that cause *X* and instead depends only on the value taken by *I*.

Let *X* be a variable with values 0,1 according as Bob gets outcome *A*, $A^\perp$ respectively, and *Y* be a variable with values 0,1 according as Alice gets outcome *A*, $A^\perp$ respectively for their space-like separated measurements of polarization along axis *a* in state $|\Phi^+\rangle$: and let *J* be some hypothetical intervention variable for *X* with respect to *Y* here. The counterfactual conditional (*B*) holds (at least for Bob-at-$t_2$), and we are assuming that *J* is an intervention variable for *X* with respect to *Y*. So we may conclude that the value of *X* causes the corresponding value of *Y*. By parallel reasoning applied to (*A*) we may conclude that *Y* causes the corresponding value of *X*. But this is now in contradiction to (I2). By *reductio ad absurdum*, *J* is not an intervention variable for *X* with respect to *Y*. But *J* was arbitrary. Hence there can be no intervention variable for *X* with respect to *Y*. Therefore the counterfactuals (*A*), (*B*) are not causal. Maudlin's sufficient condition for a causal connection is not correct, as applied to Alice and Bob's outcomes in this case.

  The preceding argument depended on a particular implementation of the thought that a causal connection provides a potential means of manipulating an event by intervening on its causes. There are those who will continue to reject either the thought or its implementation, perhaps because of a commitment to a metaphysical view according to which

> ...causality consist in the derivativeness of an effect from its causes. This is the core, the common feature, of causality in its various kinds. Effects derive from, arise out of, come of, their causes. (Anscombe [1971])

Their alternative thought may be that effects simply emerge from their causes as the world evolves, whether or not we choose to regard some of these events as potential loci of intervention, by agents or anything else. It is clear that quantum theory does not encourage this thought, as presently interpreted. But since it does still give agents a reason to endorse counterfactuals like (*A*) and (*B*), what can be their function, if not to point to causal connections in nature?

  To begin to answer this question, return to the information-maximizing role of counterfactual claims in an agent's theoretical reasoning. An agent is often not sure whether some event *e* occurs (read tenselessly), and is presently in a position neither to find out whether it occurs nor to bring about its occurrence. But if that event *does* occur, then her present information may permit her to infer (or estimate the probability) that some other events about which she is equally ignorant occur. So to maximize the value of the information she has in anticipating how that information may grow or change, she needs to indulge in hypothetical reasoning based on the assumption of *e*'s occurrence. If she is not sure which of a range of strictly alternative events $\{e_i\}$ occurs, permitting inferences to incompatible conclusions about the occurrence of other events (say $\{f_i\}$), then the permitted inferences may be cast in the form of subjunctive conditionals $e_i \,\square\!\!\rightarrow f_i$, all but one of which have false antecedents.

  Bob-at-$t_1$ is in just this position. Suppose he knows that Alice measures *a*-polarization of one photon of a pair in state $|\Phi^+\rangle$ at 1, and intends to measure *a*-polarization of the other photon at 2. Although he is then inevitably ignorant of the outcomes of both measurements, he can use quantum theory to form the firm expectation that they will be the same, and express this conclusion in the form of two subjunctive conditionals:

> If my outcome were to be *A*, then Alice's outcome would be *A*.
> If my outcome were to be $A^\perp$ then Alice's outcome would be $A^\perp$.

Together these subjunctive conditionals permit him to infer that, whatever his actual outcome, had that outcome been different then the outcome of Alice's measurement would have been different, even though everything in the past light cone of 1 was the same, which is (*B*).

  This does not provide (*B*) with a direct role in enhancing Bob's information: at best, it helps shape his *attitudes* at $t_2$ or later toward events about which he is already fully informed. (*B*) may be considered a by-product of the subjunctive conditionals highlighted in the previous paragraph, and these did have a direct role in enhancing Bob's conditional information at $t_1$ concerning events at 1. Viewed in this light, (*B*) serves to diagnose informational utility rather than to provide it. Causal counterfactuals are similarly diagnostic. It does not help to know that if I had not turned the key the engine would not have started if this gives me no reason to think that were I to turn the key the engine would start. While (*B*) is a symptom of Bob's potential for informational enrichment, (*A*) but not (*B*) is a similar symptom for Alice. Different subjunctive conditionals are epistemically useful to Alice and Bob because each has access to different information, given their different physical (specifically spatiotemporal) situations.

To sum up, what Maudlin presents as a sufficient condition for superluminal causal influence is not that at all, but rather a condition for diagnosing valuable epistemic links between space-like separated events revealed by quantum theory.

While conditionals such as (*A*), (*B*), (10) mark dependence relations between events at separate space-time locations, this dependence is not causal. These subjunctive conditionals serve to mark the *dependable* nature of the correlations to which they correspond that enables them to serve the epistemic purposes of localized agents. Quantum-endorsed counterfactuals are not evidence of agent-independent causal connections in nature: they are symptoms of informational resources available to physically situated agents in our world. The next section shows how the quantum states and probabilities that support them are ideally suited to meet the informational needs of physically situated agents in a relativistic world.

## 8. Counterfactuals, information and chance in a relativistic world

Having shown to his own satisfaction that the predictions of any "serious candidate theory" obeying his condition of local causality are in conflict with experimentally confirmed predictions of quantum theory, Bell took this as a symptom of a serious problem for quantum theory itself.

> For me then this is the real problem with quantum theory: the apparently essential conflict between any sharp formulation and fundamental relativity. That is to say, we have an apparent incompatibility, at the deepest level, between the two fundamental pillars of contemporary theory... ([2004], p. 172.)

His fellow travelers have endorsed this opinion (Maudlin [2011], Norsen [2011], Seevinck [2010]) even while some have expressed hope for a solution in work such as that of Tumulka ([2006], [2009]).

I shall argue against this consensus that, when interpreted along the lines of section 3, quantum theory fits comfortably within relativistic space-time. Indeed, quantum probabilities and the informational resources they bring are wonderfully adapted to enable agents situated in relativistic space-time to exploit the possibilities inherent in the light cone structure that is so central to relativity.

How can an agent localized in space-time obtain information about what is happening elsewhere? There are two obvious methods open to her, on the assumption that no physical process propagates superluminally. She can perform observations and experiments that access information in her backward light cone *via* physical processes capable of conveying information to her from her (absolute) past: if she is patient, she can wait for more such information to become accessible to her as her world-line extends into her future. Or she can seek knowledge in intention of future events by acting so as make things happen in her forward light cone. Neither method is infallible—but no pragmatist would expect infallibility.

There are also less obvious methods. She can extrapolate robust regularities ("laws of nature") evidenced locally to regions of space-time, even outside her light cone, as cosmologists do when drawing reasonable conclusions about happenings outside our particle horizon. More familiarly, she can appeal to correlations between joint effects of a common cause in her backward light cone, at least one of which is locally accessible, to make reasonable inferences to distant events—again including those outside her light cone. For example, two agents Alice and Bob can meet somewhere on earth and agree both to eat chocolate ice cream at some future time *t* if the random number each receives a minute earlier broadcast from a radio transmitter on a satellite orbiting Venus is even, but strawberry if it is odd. Bob travels sedately to Mars, which is in conjunction to the sun at the appointed hour. If all goes well, Alice knows what flavor of ice cream Bob eats on Mars at space-like separation from her at *t*.

As is now well known to quantum information theorists, "non-local" quantum correlations permit wholly novel methods of acquiring and sharing information. Here is an example to be used to illustrate points to come, based on the correlations inherent in the GHZ state (cf. Ghirardi [2005]). Alice, Bob and Charlie are set a coordination task, to be undertaken when each is at a location space-like separated from the others. When together beforehand, they are told that a fair coin will be tossed at each location, after which each will be asked to write down either the number +1 or the number −1. Their job is to devise a strategy and to implement it to ensure that
(i) If all coins land heads, then the product of the numbers they have written down is −1.
(ii) If exactly one coin lands heads, then the product of the numbers they have written down is +1.
(What happens in the other four possible combinations of outcomes of coin tosses doesn't matter.)
Assuming no superluminal communication, none of them can know either the outcome of the others' coin tosses or what they have written down before writing down his or her own number.

Any strategy they devise before separating that relies at most on correlations predictable within classical physics (with no superluminal physical processes) can be modeled as a choice of the variable λ in the argument of CPP sketched in the last section. That argument is readily adapted to show that Alice, Bob and Charlie have no successful strategy of that kind. But the "non-local" correlations predicted in quantum theory by application of the Born Rule to the GHZ state *do* make available a successful strategy. Each of A, B, C should measure a polarization component of his or her photon from a GHZ state they jointly prepared before separating. If his or her coin landed heads, this should be linear polarization along /, and he or she should write down +1 if the photon is found to be polarized along that direction, otherwise −1. If his or her coin landed heads, this should be circular polarization, and he or she should write down +1 if the photon is found to be right-circularly polarized, otherwise −1. Setting aside practical implementation problems, quantum theory assures each of them that this strategy will (with probability 1) succeed.

The GHZ correlations do not tell Alice (say) either how Bob's and Charlie's coins land or what number either of them writes down, even given the outcomes of her own coin toss and chosen polarization measurement. But knowledge of the strategy based on them does supply her with *conditional* information about what happens at Bob and Charlie's locations. If her coin lands heads and she finds her photon polarized along axis /, then she can infer that, were Bob and Charlie's coins both to land heads, then one of them (she can't say which) would write down +1 while the other would write down −1: while if both of Bob and Charlie's coins were to land tails then they would each write down the same number (she can't say whether this is +1 or −1). If, on the other hand, she finds her photon polarized orthogonal to axis / then she can infer that, if Bob and Charlie's coins were to land heads, then they would write down the same number: while if both of Bob and Charlie's coins were to land tails then one of them (she can't say which) would write down +1 while the other would write down −1.

Such conditional information is just what each player needs to be sure their strategy will succeed. It is subjunctive conditionals endorsed by quantum theory (such as (10)) that supply them with this information. That is how the "non-local" correlations predicted by quantum theory enhance the informational resources available to agents who accept it in a way that has no parallel in classical physics. The enhancement depends on endorsement of the subjunctive conditionals supported by those correlations. That is why it is important for agents using quantum theory to accept these conditionals, and the counterfactual conditionals (formed by denying their antecedents) that are diagnostic for them.

Notice that nothing about the application of GHZ correlations in the three preceding paragraphs depended on the space-time intervals separating A,B,C as each implements the successful strategy. The subjunctive conditionals on which its success depends, and the Born probabilities that underlie them, are insensitive to where in space-time each is when implementing the strategy. So while the success of the strategy when these intervals are all space-like cannot be emulated classically, the informational resources inherent in quantum correlations such as these can be deployed to advantage in situations where a classical emulation is available. Here are two further examples.

The Bell state $|\Phi^+\rangle$ predicts perfect (probability1) correlations for a measurement of H/V linear polarization on each of two photons by Alice and Bob irrespective of the space-time interval separating them. Knowing the outcome of her own polarization measurement, Alice can use one of two subjunctive conditionals to infer the outcome of Bob's polarization measurement.

If A's outcome were to be H, then B's outcome would be H.
If A's outcome were to be V, then B's outcome would be V.

If their measurements are space-like separated (as in figure 3), these outcomes do not result from a common cause $\lambda$ in the overlap of their backward light cones satisfying equations (5*a*), (5*b*). Suppose that Alice and Bob had wished to arrange beforehand secretly to coordinate their activities in regions 1,2 respectively. They could have done so by means of a classical process serving as a common cause of events in those regions simply by carrying shared private instructions to 1, 2 respectively. But then information about events in region 2 would have been available in the backward light cone of 1 (as also of 2) to serve as the source of Alice's knowledge about region 2. However, if they arrange each to measure H/V polarization on state $|\Phi^+\rangle$, they can coordinate their activities in 1, 2 while leaving nothing in the backward light cone of 1 or 2 to serve as a source of information to a potential "spy" as to what they will do. In this way, quantum correlations permit an agent to acquire essentially *private* information about happenings space-like separated from her. This is just one simple example of the rich cryptographic potential inherent in subjunctive conditionals endorsed by quantum theory.

Now consider the entanglement-swapping scenario depicted in figure 5 in which the polarizations of all four photons in two entangled photon pairs, each independently prepared in Bell state $|\Psi^-\rangle = 1/\sqrt{2}\,(|HV\rangle - |VH\rangle)$, are measured in a four-fold coincidence.[13] When the Bell-state analyzer is switched to the appropriate setting, Victor measures the polarization of any photons emerging to the left and right of the beam splitter on which photons 2 and 3 are simultaneously incident. If he wishes to use quantum theory to guide his expectations as to the outcomes of Alice's and Bob's polarization measurements, then what entangled polarization state he should assign to the composite system composed of photons 1 and 4 prior to detection depends on the outcomes of his own measurements, which may be arranged either to be space-like separated from Alice and Bob's measurements or to lie in the overlap of their forward light cones.

In the former case, he can apply quantum theory to his outcomes to generate subjunctive conditionals (some strict, others probabilistic) concerning the outcomes of whatever polarization components Alice and Bob should choose to measure. No classical process within his light cone could provide him with this information.

However, in case Alice and Bob's measurements occur in Victor's backward light cone, then of course there are subluminal classical processes that Victor might be able to exploit to fully inform himself of Alice and Bob's measurements and their outcomes. But Victor can avail himself of these only if the necessary physical connections are actually in place. There are important lessons here for the nature of quantum states and chances in a relativistic world.

Suppose that in an ideal experiment of this form Alice and Bob each measure the polarization of their photon using a highly efficient birefringent crystal with a detector in only one output channel. In these circumstances, failure to detect a photon indicates presence of a photon in the other channel. If suitable processes occur within Victor's backward light cone, then with careful timing he can use them to inform himself of cases in which both Alice and Bob have failed to detect photons in coincidence with the two photons he has detected. Suppose that in one such case Victor applies quantum theory to his own outcomes to assign some entangled polarization state to the composite system composed of photons 1 and 4 immediately before Alice and Bob's measurements. That quantum state does not represent any intrinsic property of either photon, or even of the pair as a whole: as presently understood, a quantum state never represents an intrinsic property of a system to which it is ascribed. But nor, in this situation does it represent Victor's best source of advice on what to believe about Alice and Bob's measurements on photons 1 and 4 or their outcomes—he already knows all about that. So Born probabilities calculated from this entangled polarization state are "trumped" as a guide to Victor's expectations by the extremal (0 or 1) probabilities delivered through his classical channels.

If Victor is physically situated so that he actually has access to reliable information about Alice and Bob's measurement outcomes through channels provided by processes within his backward light cone, then he should use that information to guide his beliefs about these outcomes rather than relying on Born probabilities derived from the entangled polarization state he justifiably assigns to 1+4 after consulting the outcomes of his own polarization measurements. Moreover, he should then assign each of 1 and 4 after Alice and Bob's measurements (respectively) the eigenstate of linear polarization associated with the channel with no detector in it after emerging from their respective birefringent crystals.

Consider what this means for objective chance in quantum theory. The motivating idea behind Lewis's [1980] Principal Principle connecting credence to objective chance was that objective chance should be understood as the "ideal" credence an agent should adopt, given all admissible information. But, as he stated it, the Principle indexes chance to (absolute) time, as follows:

(PP) Let $C$ be any initially reasonable credence function. Let $t$ be any time. Let $x$ be any real number in the unit interval. Let $X$ be the proposition that the chance, at time $t$, of $A$'s holding equals $x$. Let $E$ be any proposition compatible with $X$ that is admissible at time $t$. Then $C(A/XE)=x$. ([1986], p.87)

Subsequent critical discussion focused on what propositions should be considered admissible, and why. Lewis's indexing admissibility to time makes his principle ill-adapted to a relativistic space-time structure. But suppose one modifies the principle as follows:

(MPP) Let $C$ be any initially reasonable credence function. Let $p$ be any space-time location of a physically situated agent. Let $x$ be any real number in the unit interval. Let $X$ be the proposition that the chance, at $p$, of $A$'s holding equals $x$. Let $E$ be any proposition compatible with $X$ that is admissible at $p$. Then $C(A/XE)=x$.

Here is one natural way to understand admissibility in this context: A proposition is admissible at $p$

just in case it is a consequence of whatever has happened in the backward light cone of *p* and what this implies for happenings outside *p*'s backward light cone. With this modification, one can ask: Can Born probabilities be understood as modified Lewisian chances?

Quantum theory is a fallible product of our scientific efforts. If one thinks of chance as an objective feature of the world, then to specify a Born probability is at best to attempt correctly to state a chance. From another perspective, chance constitutes an ideal credence at which an agent should aim when using all available information, including her best scientific theories. In the light of relativity, we (accepting relativity) may take an agent physically situated at *p* ideally to have available complete information as to what has happened in the backward light cone of *p*, together with whatever she may infer from this using quantum theory (which we take it she should accept) together with the rest of our science. Knowing everything about the contents of her backward light cone, she may be supposed to have information sufficient to assign quantum states to various systems, and to use these assignments to calculate Born probabilities for actual and merely hypothetical measurement outcomes. We suppose that sometimes such assignments give an agent's ideal credences at *p*. Alice–at–$t_2$'s Born probabilities in figure 3 are her ideal credences concerning the outcome of her measurement in region 1, or so we believe, while Bob–at–$t_2$'s Born probabilities are his ideal credences concerning the outcome of Alice's measurement in region 1. We believe this because we believe Alice–at–$t_2$ and Bob–at–$t_2$ have each made the best possible use of all information available to them in assigning their respective credences, even though these differ. So we may take these different Born probabilities as chances subject to (MPP). Then it becomes clear why Bell's local causality condition is not appropriately applied to theories that specify objective chances in a relativistic world.

The case is different for Victor in figure 5. Suppose Victor assigns an entangled polarization state to 1+4 in the light of his own polarization measurements and uses this to calculate Born probabilities for Alice and Bob's measurement outcomes, both in his own backward light cone. As a practical matter, this may give him his best way of assigning credences to these various outcomes (perhaps in combination with the credences he attaches to each different measurement he supposes Alice and Bob may have made). But ideally he has available to him complete information about what has happened in his backward light cone, and this would enable him to assign credence 0 or 1 to Alice and Bob's outcomes. Victor's Born probabilities for Alice and Bob's measurement outcomes do *not* correspond to his ideal credences in this case, so these Born probabilities are not objective chances subject to (MPP)—or so we believe.

## 9. Conclusion
It is time to remove the scare quotes around the word 'non-local' in the title of this paper by saying in what sense patterns of statistical correlation among distant events, first predicted using quantum theory and now amply confirmed by experiment, count as non-local.

The correlations that display these patterns are not *localized* when they are manifested by sets of distinct-event *n*-tuples (*n*=2,3, ...) in which at least two distinct events occur at different locations. This could happen in one of two ways. While space-like (or null) separated events occur at different locations in all reference frames, time-like separated events may occur at different locations in some salient reference frame (e.g. that of the laboratory). The mere occurrence of patterns of non-localized correlations has no implications for causation or for relativity theory. To derive such implications it is necessary to assume some theoretical explanation of the observed patterns. But whether that proves sufficient will depend on the theory, and on how it helps one explain these patterns.

Bell used an intuitive principle of local causality to motivate the imposition of a local causality condition on any theory of a certain form, advanced as offering the possibility of explaining patterns of non-localized correlations successfully predicted by quantum theory. Any theory of that form is able sufficiently to represent the contents of the backward light cone of a region 1 in which an event in an *n*-tuple may occur by values taken there by magnitudes he calls local beables—sufficient, that is, to determine a unique probability attached to values of magnitudes in 1 determining the occurrence such an event. His local causality condition then requires that this unique probability be unaltered by specification of values taken by magnitudes in any region space-like separated from 1. He shows that no theory of that form meeting this condition can explain all the observed patterns of correlation. This is the precise sense emerging from Bell's work in which the observed patterns of correlation are not locally causal. If that exhausts the content of the claim that these correlations are non-local the claim is strictly correct. But no-one hearing that claim is likely to understand its content to be so restricted, so the claim is then liable seriously to mislead.

Bell showed that no theory of the right form meeting his condition of local causality can explain all the observed patterns of correlation by deriving a CHSH inequality these patterns violate. It is becoming increasingly common to call the observed patterns themselves non-local simply because they violate an inequality like that first derived by CHSH [1969] and Bell ([2004], p.37, p.57). This usage has been extended to apply to what are called 'non-local boxes'—hypothetical devices that would not permit signaling but output statistics violating a CHSH or other inequality, perhaps by even more than do the quantum statistics. This extension is useful for certain purposes. But it is important to reiterate the fact that such actual or hypothetical statistics by themselves have no implications for causation or relativity.[14]

Quantum theory does not take a form suitable for application of Bell's condition of local causality. So while it cannot be said to be locally causal, nor does it *fail* to be locally causal. Application of the Born Rule does not yield a unique probability for an event in an *n*-tuple to occur in region 1, but different probabilities relative to different space-time locations that may (or may not) be occupied by agents well-advised to base their credences on it concerning events in 1. Agents basing their expectations on this advice will (almost certainly) record statistics conforming to the patterns quantum theory leads them to expect—the patterns of non-localized correlations that no theory meeting Bell's condition of local causality can explain. That is how quantum theory helps them explain EPR-Bell correlations displayed by pairs of quantum systems, GHZ correlations displayed by triples, and so on. That is how quantum theory helps us explain "non-local" correlations—i.e. statistical patterns of non-localized correlations. The explanation involves no causal relations, neither local nor non-local, so it conforms vacuously to Bell's intuitive principle of local causality. It is local in the sense that any actual or potential localized agent can apply quantum theory to show that anyone in her spatiotemporal position should expect to observe just what she does in fact observe. Section 6 did not discuss the Lorentz transformation properties of the quantum states assigned, relative to different space-time regions, in explaining observed patterns of correlations. But in relativistic quantum theory these states transform in the usual way under Lorentz transformations. Both quantum theory and expectations based on it are not merely compatible with relativity, but enhance the abilities of agents situated in relativistic space-time to exploit the informational capacities of their world.

**Acknowledgments**
This publication was made possible through the support of a grant from the John Templeton Foundation. The opinions expressed in this publication are those of the author and do not necessarily reflect the views of the John Templeton Foundation. I thank Prof. A. Zeilinger and Prof. M. Aspelmeyer, my hosts at the Institute for Quantum Optics and Quantum Information of the Austrian Academy of Sciences under the Templeton Research Fellows Program "Philosophers-Physicists Cooperation Project on the Nature of Quantum Reality", and the other scientists in the Institute for many enlightening discussions. I acknowledge the support of the Perimeter Institute for Theoretical Physics under their sabbatical program during my visit in Fall 2009: research at the Perimeter Institute is supported by the Government of Canada through Industry Canada and by the Province of Ontario through the Ministry of Research and Innovation.

Sussex: Wiley-Blackwell.
Norsen, T. [2009]: "Local Causality and Completeness: Bell *vs.* Jarrett", *Foundations of Physics 39*, pp.273-294.
-------------[2011]: "J.S. Bell's Concept of Local Causality",
http://arxiv.org/PS_cache/arxiv/pdf/0707/0707.0401v3.pdf
Pan, J.-W., Bouwmeester, D., Daniell, M., Weinfurter, H., Zeilinger, A. [2000]: "Experimental test of quantum non-locality in three-photon Greenberger-Horne-Zeilinger entanglement", *Nature 403*, 515-9.
Pearle, J. [2009]: *Causality*. Cambridge: Cambridge University Press.
Peres, A. [1978]: "Unperformed experiments have no results", *American Journal of Physics, 46*, 745-7.
Reichenbach, H. [1956]: *The Direction of Time*, Berkeley, University of Los Angeles Press.
Seevinck, M.P. [2010]: "Can quantum theory and special relativity peacefully coexist?", http://arxiv.org/abs/1010.3714
Seevinck, M.P., Uffink, J. [2010]: "Not throwing out the baby with the bathwater: Bell's condition of local causality mathematically 'sharp and clean'", http://arxiv.org/abs/1007.3724
Tumulka, R. [2006]: "A Relativistic Version of the Ghirardi-Rimini-Weber Model", *Journal of Statistical Physics, 125*, 821-40.
---------------[2009]: "The Point Processes of the GRW Theory of Wave Function Collapse", *Reviews in Mathematical Physics, 21*, 155-227.
Woodward, J. [2003]: *Making Things Happen*. Oxford: Oxford University Press.

**Figures**

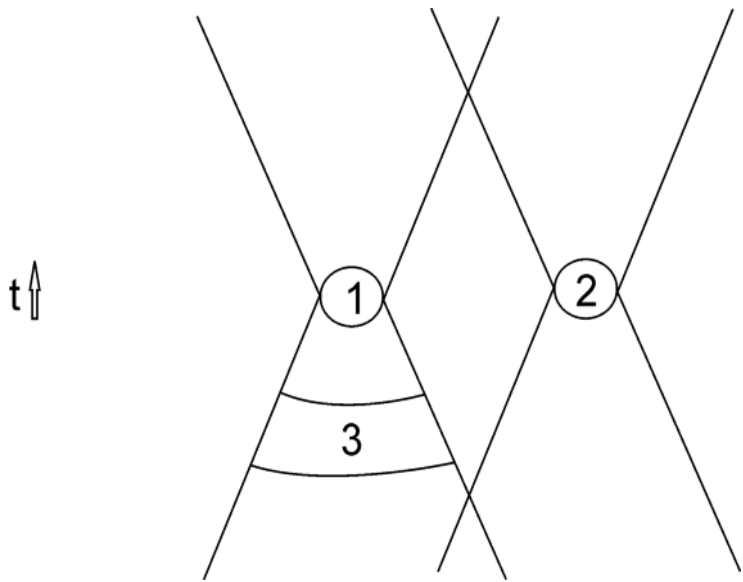

Figure 1

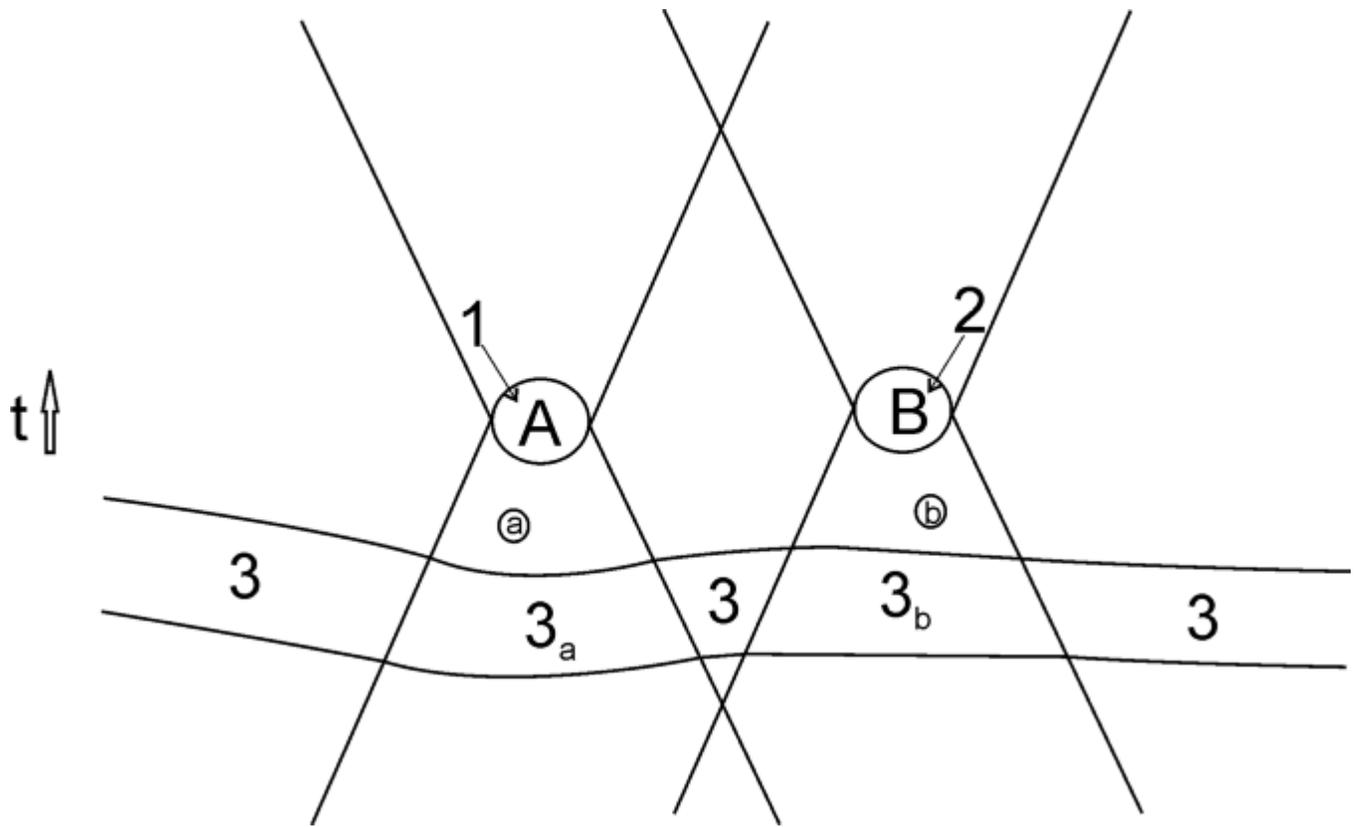

Figure 2

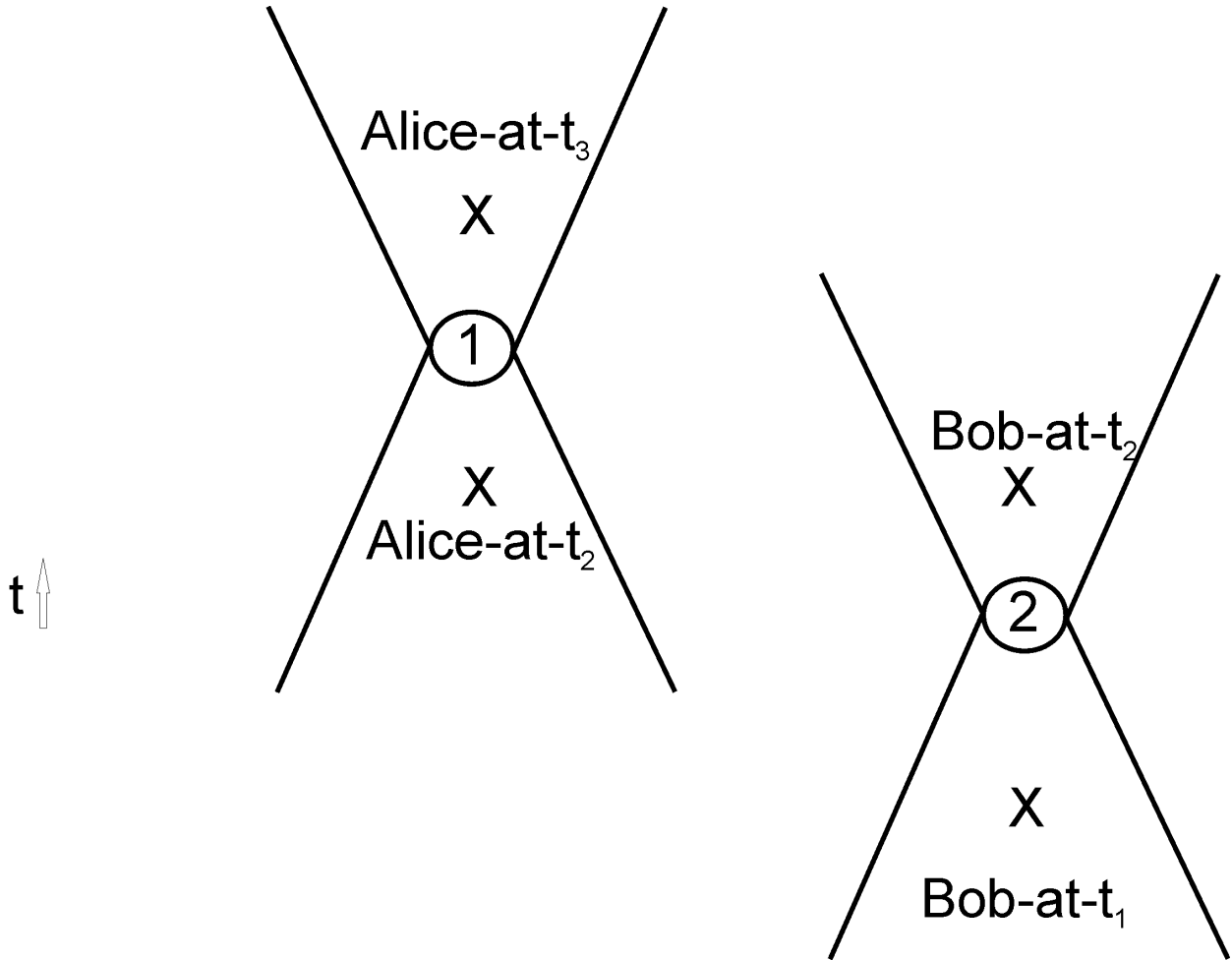

Figure 3

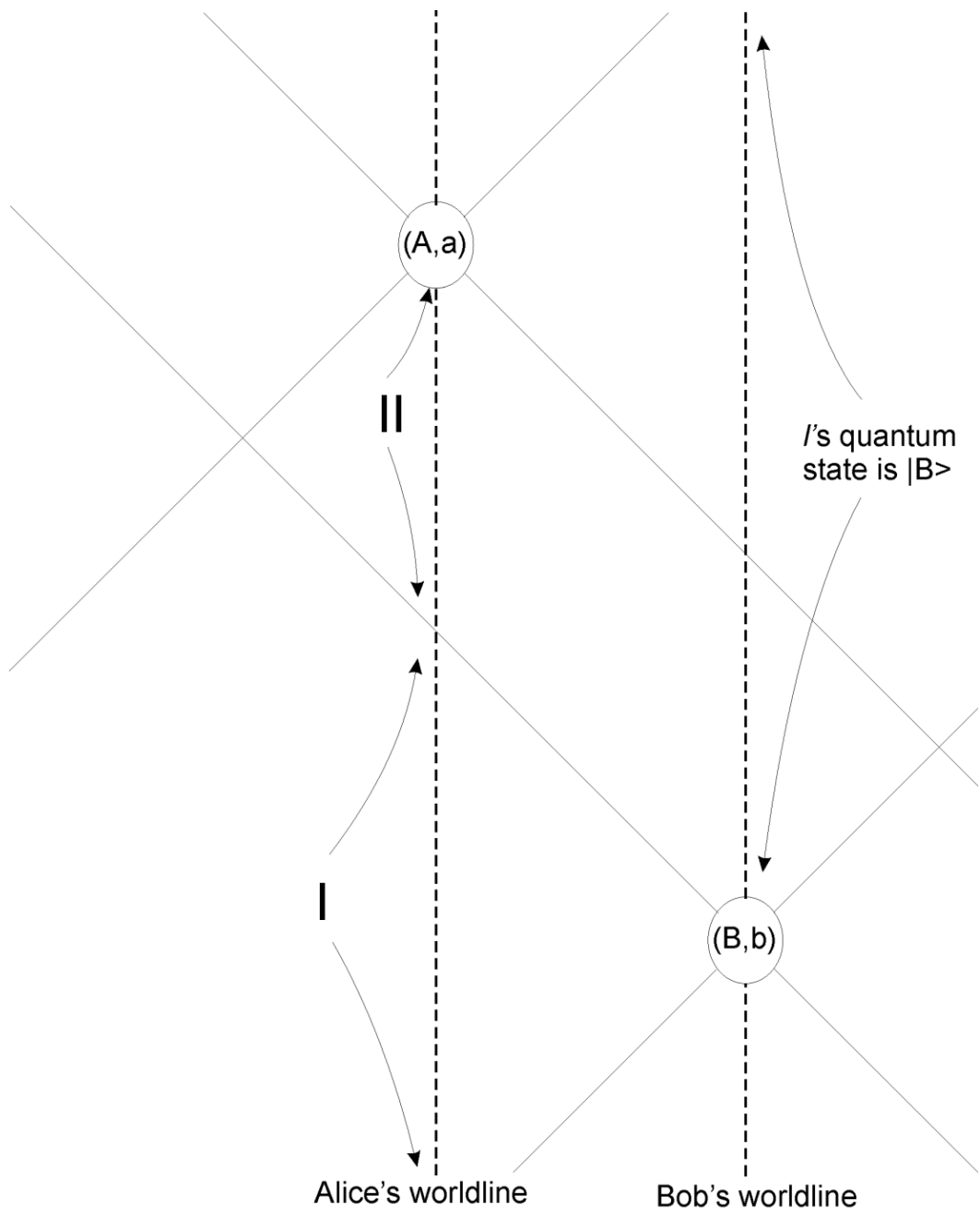

Figure 4

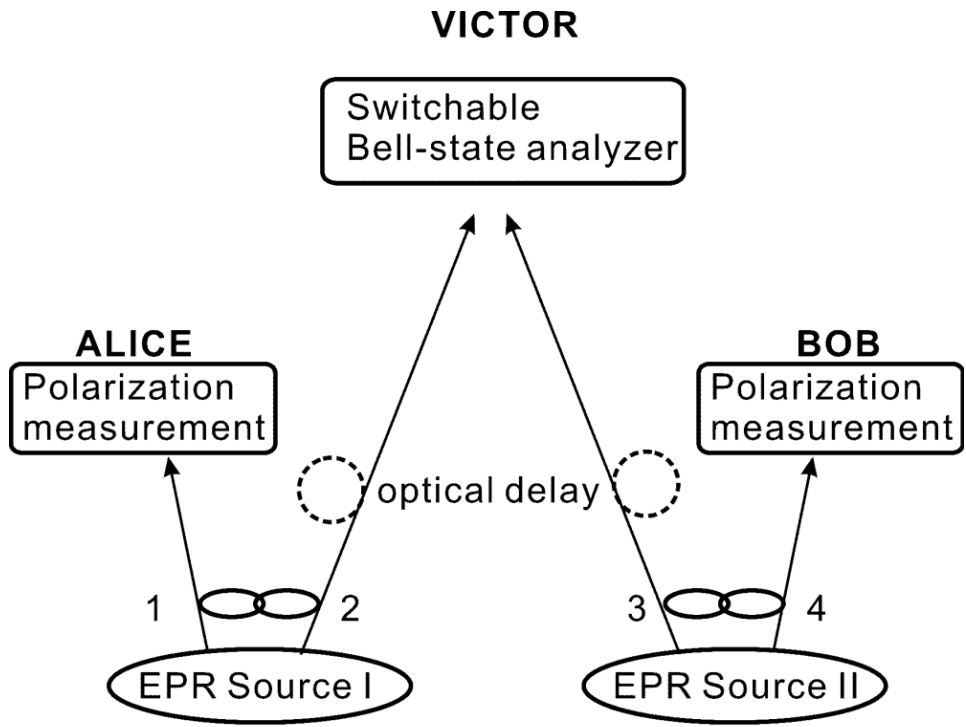

Figure 5

1. Especially Norsen ( [2009], [2011]), Seevinck and Uffink [2010], Goldstein, S. *et*. a*l*. [2011].

2. ($\mathbf{P}^Q[\Delta]$ is an element of the projection-valued measure defined by the self-adjoint operator $Q$ representing $Q$, and Tr is the trace operation).

3. There is certainly room to doubt that Bohr held either assumption, and Bohr's Copenhagen interpretation surely lays claim to orthodoxy.

4. This makes it clear that a claim should be thought of as a meaningful assertion rather than a proposition asserted. One could try to model a NQMC as expressing a determinate proposition whenever made, though a different proposition in different contexts. But it is not clear how useful it will be to introduce propositions into an inferentialist semantics. To resolve this issue it will be necessary to develop elements of such a semantics and to apply it to the use of NQMC's in the light of quantum theory,

5. Bell's 1975 presentation of local causality (reprinted in Bell [2004], pp. 52-62) took this as a generalization of local determinism in a theory. While a specification of the local beables of a locally deterministic theory (such as the fields in Maxwell's electromagnetism) in the backward light cone of a region wholly determine the local beables in that region, Bell took it that the local beables of a *stochastic* theory in the backward light cone of a region wholly determine the *probabilities* of local beables in that region. On that understanding, local causality is indeed a natural generalization of local determinism: uniquely determined *values* for local beables are simply generalized to uniquely determined *probabilities* for values of local beables. The subsequent discussion has largely followed Bell here in taking it for granted that any theory that is a serious candidate for explaining EPR-Bell correlations must be either deterministic (uniquely specifying values of local beables) or stochastic (uniquely specifying their *probabilities*, which accordingly themselves come to be thought of as values of local beables).

6. Note that any apparent overall temporal asymmetry between Alice and Bob suggested by the diagram is an artefact of the choice of inertial frame that makes Alice-at-$t_2$ simultaneous with Bob-at-$t_2$: a different choice would make Alice-at-$t_3$ simultaneous with Bob-at-$t_1$. Of course, in ordinary circumstances the salient choice would be their shared laboratory frame.

7. What makes a physical situation relevant here is being in the forward light cone of neither 1 nor 2 and having potential access to information justifying a claim that the state in 3 is $|\Phi^+\rangle$.

8. It violates common cause principles like that of Reichenbach [1956]. Arntzenius [1999] surveys several such conditions and their limitations, attributing their hold on us to our physical and practical situation as agents in a thermodynamically temporally-asymmetric world.

9. Peres [1978] registered his dissent from this consensus with the memorable remark "Unperformed experiments have no results." Stairs [forthcoming] attacks what he calls the EPR illusion that associates the truth of measurement counterfactuals with "non-local" correlations.

10. CPP appeal to this formulation of Lewisian strict causal dependence:

> If $c_1, c_2, ...$ and $e_1, e_2, ...$ are two pairwise distinct families of events (i.e. $c_1$ is distinct from $e_1$, $c_2$ is distinct from $e_2$, etc.) such that no two of the $c$'s and no two of the $e$'s are compossible, then the '$e$-family' deterministically causally depends on the '$c$-family' at world $w$ if and only if $c_1 \,\square\!\!\rightarrow e_1$, $c_2 \,\square\!\!\rightarrow e_2$, etc. are true at $w$.

In their argument they apply it to a case in which each $c_i$ is composed of several sub-events $c_i^1$, $c_i^2$, etc. that act as partial causes of $e_1$ to show that Lewisian causation obtains between an event and some of its partial causes at space-like separation.

11. See in particular Pearle [2009], Woodward [2003].

12. In earlier work including my [1992], [1994], I took the possibility of manipulating one event by intervening in another as just one element of a cluster concept of causation, and held that a usefully revised concept of causation can survive its removal from the cluster. But that non-interventionist concept supported a notion of "causal" explanation of Alice's outcome in terms of Bob's (and vice versa) only in so far as quantum theory can be taken to describe what in my [1994] I called a non-separable process linking these outcomes. Interpreted along the lines of section 3 of the present paper, quantum theory describes no such process. So one cannot appeal to this non-interventionist revision of our causal concept in arguing that counterfactuals like (*A*), (*B*) have causal implications, even in this revised sense of 'causal'.

13. For further details, see my [in press].

14. While the condition that a non-local box not permit signaling may seem to have implications for causation, it simply requires that the marginal probability distribution for each output be independent of all other inputs.